\documentclass[amsmath,amssymb,showpacs]{revtex4}
\usepackage{amsmath,amsfonts,amssymb,latexsym,graphicx,youngtab}
\usepackage{tikz}
\newtheorem{thm}{Theorem}
\newtheorem{lemma}{Lemma}

\newtheorem{rema}{Remark}

\renewcommand{\theequation}{\arabic{equation}}
\renewcommand{\theequation}{\arabic{equation}}
\renewcommand{\thethm}{\arabic{thm}$\:$}
\renewcommand{\therema}{{\em{\bf \arabic{rema}$\:$}}}
\renewcommand{\thelemma}{\arabic{lemma}$\:$}

\newcommand{\dis}{\displaystyle}

\newcommand{\bequ}{\begin{equation}}
\newcommand{\eequ}{\end{equation}}
\newcommand{\barr}{\begin{array}}
\newcommand{\earr}{\end{array}}
\newcommand{\bea}{\begin {eqnarray}}
\newcommand{\eea}{\end {eqnarray}}
\newcommand{\lb}{\label}

\renewcommand{\Im}{{\cal I}{\rm m}\:}
\renewcommand{\Re}{{\cal R}{\rm e}\:}
\begin{document}
\def \tr {\mathrm{Tr\,}}
\let\la=\lambda
\def \Z {\mathbb{Z}}
\def \Zt {\mathbb{Z}_o^4}
\def \R {\mathbb{R}}
\def \C {\mathbb{C}}
\def \La {\Lambda}
\def \ka {\kappa}
\def \vphi {\varphi}
\def \Zd {\Z ^d}
\title{On Charge Conjugation, Correlations, Elitzur´s Theorem and the Mass Gap Problem in Lattice $\mathrm{SU}(N)$ Yang-Mills Models in $d=4$  Dimensions}
\author{Paulo A. Faria da Veiga} \email
{veiga@icmc.usp.br}
\author{Michael O'Carroll}\email{michaelocarroll@gmail.com}
\affiliation{Departamento de Matem\'atica Aplicada e
Estat\'{\i}stica, ICMC-USP,
\\
Av. Trabalhador S\~ao Carlense 400, 13566-590 S\~ao Carlos SP, Brazil}

\date{January 13, 2026\vspace{.3cm}}
\begin{abstract}
We consider a $d=4$ spacetime Euclidean dimensional Wilson lattice Yang-Mills model with gauge group ${\mathrm SU}(N)$, and the associated lattice Euclidean quantum field theory constructed by Osterwalder-Schrader-Seiler via a Feynman-Kac formula. In the Wilson model, to each lattice bond $b$ there is assigned a bond variable $U_b\in{\mathrm SU}(N)$. The gluon fields are parameters in the Lie algebra of ${\mathrm SU}(N)$. We define a charge conjugation operator $\mathcal C$ in the underlying physical quantum mechanical Hilbert space $\mathcal H$ and prove that, for $N\not= 2$, $\mathcal H$ admits an orthogonal decomposition into two sectors with charge conjugation $\pm 1$. There is only one sector for $N=2$. In the space of correlations, a charge conjugation operator ${\mathcal C}_E$ is defined and an analogous decomposition holds. Besides, a version of Elitzur´s theorem is shown and applications are given. It is proven that the expectation averages of two distinct lattice vector potential correlators is zero. Surprisingly, the expectation of two distinct field strength tensors is also zero. It is known that, in the gluon field parametrization, the Wilson action is bounded (from above!) quadratically in the gluon fields. Towards solving the mass gap problem, in the gluon field expansion of the action, we show that there are no local terms, such as a mass term. However, the action associated with the exponent of the exponentiated Haar measure density has a positive local mass term which is proportional to the square of the gauge coupling $\beta = g^{-2} >0$ and the ${\mathrm SU}(N)$ quadratic Casimir operator eigenvalue. Although this property does not prove the mass gap, it indicates a mass gap is developed. Precisely, for Yang-Mills models, the spectral mass gap is associated with the exponential distance decay of the truncated two (gauge-invariant) plaquette field correlations via the Payley-Wiener theorem. Our results also hold $d=3$. The orthogonal decomposition of $\mathcal H$ has consequences in the analysis of the truncated two-point plaquette field correlation and the Yang-Mills mass gap problem: for $N\not=2$ and for, at least, for small  $\beta$, a multiplicity-two one-particle glueball state (determining two mass gaps)  is expected to be present in the low-lying energy-momentum spectrum of the Yang-Mills model.
\end{abstract}
\maketitle
\section{INTRODUCTION} \lb{sec1}
Nonabelian local gauge-invariant Yang-Mills (YM) models were introduced long ago in \cite{YMOld}. Later, the $\mathrm{SU}(3)$ model acquired a major relevancy in Quantum Chromodynamics (QCD) as the best candidate for the description of strong interactions  Ref. \cite{FGL}. 

K. Wilson (see Refs. \cite{Wil2,Wil} and references therein) introduced the lattice approximation to the continuum spacetime, which provides an ultraviolet cutoff for the model, and developed a functional integral formalism for QCD on the lattice. This approach was powerful enough to e.g. obtain the first results on the QCD particle spectrum and to exhibit confinement. Many applications came soon after either within the strong coupling expansion or in other contexts (see e.g. Refs.
\cite{Wil,Banks,FK,Mon,MW,Schreiber,Creu,Mach,Mach2} and the books
given in Refs. \cite{Creu2,MM}). 

In the seventies, prospecting in a complementary direction, the
works of E. Nelson, K. Symanzik, and the joint work of K.
Osterwalder, R. Schrader and E. Seiler set up the rigorous basis for the the imaginary-time functional integral formalism for models with or without fermions (see Ref. \cite{GJ} and references therein). J. Glimm, A. Jaffe and, especially, T. Spencer \cite{Sp,Si} with his hyperplane decoupling method made it possible to determine   (at least) the low-lying particle structure (spectrum) of quantum field models. (For more details about all this mathematical work, see the books \cite{GJ,Sei,Si} and references therein.) In this way, one was able to control the infinite-volume limit and to establish the gap and the upper gap properties for simple scalar models. K. Osterwalder and E. Seiler in Refs. \cite{OS,Sei} provided the formalism to extend this analysis to models with fermions. More specifically, physical positivity was established, the physical Hilbert space and the energy-momentum operators were constructed and Feynman-Kac (F-K)
formulae were obtained for correlations in lattice QCD models.
In the context of the strong coupling regime of lattice QCD, the low-lying one-particle energy-momentum hadronic spectrum and the related two-particle bound states was obtained in Refs. \cite{QCD,CMP,JMP,2baryon,2meson,2flavor2baryon,2flavor2meson,physlettb,longo,8fdw,8fdwm,PMJ1,PMJ2}.

At the level of numerical simulations, the particle description and
other important questions could be analyzed which were not attainable using the traditional
perturbation theory. For instance, in Refs. \cite{num1,num2,num3,num4} we find treatments analyzing the hadron spectrum and Refs. 
\cite{Ochs,Munster,Teper,Seo} deal   with the glueball spectrum.

In recent years, perhaps due to the fact glueballs were not experimentally detected, the glueball spectrum lost a bit of interest from the community but the existence of glueballs, the mass gap problem, and a proof of confinement manifestation in Y-M models is still one of the key problems in particle physics, together with the mathematical existence of the UV limit of QCD. The glueball mass gap problem is one of the problems taking part in the list of one-million dollar problems of the Clay Foundation.

The discrete symmetry of charge conjugation plays an important role in 
the description of glueball states. Despite of this status, we never found a 
satisfactory treatment of this symmetry and its consequence on 
the underlying quantum mechanical Hilbert space $\mathcal H$ in the literature.

Based on this fact, in this paper, we take $0< \beta\equiv(1/g^2) \ll 1$, where $g>0$ is the gauge coupling parameter in a lattice Wilson plaquette action in the Wilson formulation of YM. With this restriction on $\beta$, positivity holds and there is an underlying quantum mechanical Hilbert space ${\mathcal H}$ of physical states. Considering Euclidean spacetime dimension four (but our results extend also to dimension three) and using the Feynman-Kac formula for a half-integer shifted temporal hypercubic lattice $\Lambda\,\subset a\left[\,Z_t\,\equiv\,(\mathbb Z\,+\,\frac12)\times \mathbb{Z}^3\right]$, $a\in(0,1]$, we give a rigorous analysis for charge conjugation in lattice YM models with the gauge group  ${\mathrm SU}(N)$ and exploit its consequences involved in solving the mass gap problem.  For $N=2$ and small gauge coupling $\beta$, corresponding to the real character case, the mass gap problem was treated in Refs \cite{Schor1,Schor2,Schor3}. There is only one glueball particle in the low-lying energy-momentum spectrum. For $N\not= 2$, we show there is an orthogonal decomposition of $\mathcal H$ into two subspaces. Hence, there is a glueball state and a mass gap in each sector. 

  Moreover, we prove a version of Elitzur´s theorem, obtain the Haar measure for ${\mathrm SU}(N)$ in the gluon field parametrization and discuss the Wilson   plaquette action expansion in terms of gluon fields to show there is no appearance of a positive local mass term (a relevant term, concerning the canonical scaling in the renormalization group formalism). Taking the ${\mathrm SU}(N)$ case, we show that a local mass term arises from expanding the exponent of the exponentiated Haar measure density. Although this property does not prove the mass gap, it indicates a mass gap is developed. Precisely, for Yang-Mills models, the spectral mass gap is associated with the exponential distance decay of the truncated two (gauge-invariant) plaquette field correlations (see Ref. \cite{Schor1,Schor2,Schor3}) via the Payley-Wiener theorem \cite{Stein}.

In section \ref{sec2}, we define our lattice model, introduce the physical quantum mechanical Hilbert space $\mathcal H$, and statistical mechanical correlations via the Feynman-Kac formula. We define and determine properties of the charge conjugation operators   $\mathcal C$ and $\mathcal C_E$, respectively, in $\mathcal H$ and in the space of statistical mechanical correlations. In section \ref{sec3}, we revisit Elitzur´s theorem. Section \ref{sec4} is devoted to some concluding remarks. In Appendix A, we prove our version of Elitzur´s theorem stated in Section III and discuss gauge fixing.  In Appendix B, we analyze the Wilson action in the gluon field parametrization and the  gauge group Haar measure in the gluon field parametrization is obtained in Appendix C.
\section{The Wilson Lattice Model, Charge Conjugation and Physical Hilbert Space}\lb{sec2}
We work with the compact unitary matrix Lie gauge group $\mathcal G\,=\,{\mathrm SU}(N)$, of dimension $\delta_N\,=\,N^2-1$, and let the hypercubic lattice be $\Lambda\,\subset\,Z_t\,\equiv\,a\,[(\mathbb Z\,+\,\frac12)\,\cup\,\mathbb Z^3]$, with a lattice spacing $a\in(0,1]$. The Wilson lattice model in $\Lambda$  is defined as follows. (Our basic definitions  follow closely the treatment given in Ref. \cite{ROMP}.)

  Recalling that $\beta=(1/g^2)>0$ denotes the gauge coupling, with $g>0$, we consider the partition function for the YM lattice model in   Euclidean spacetime dimension $d=4$, with fixed $a$, gauge group $\mathcal G$ and with free boundary conditions. (With minor adaptations, our treatment can be extended to treat periodic boundary conditions and to dimension $d=3$. For $d=3$, the gauge parameter $\beta$ is the $a$-dependent function $a^{d-4}/g^2\,\equiv\,(a\beta)^{-1}$, see Eq. (\ref{partitionB}) below.) 

In the lattice $\Lambda$, the zero index labels the time direction and the spatial directions are labeled by $1,\,2,\,3$. A lattice site $x\in\Lambda$ is denoted by $x\,=\,(x^0,x^1,\ldots,x^{3})$ and the number of sites in $\Lambda$ is denoted by $\Lambda_s$. (The reason for choosing the half-integer lattice for the time coordinate is a technical one. It avoids the zero-time coordinate, so that, in the continuum limit $a\searrow 0$, two-sided equal time limits of correlations can be accommodated.) The vectors $e^\mu$, $\mu=0,1,\ldots,3$, are the unit vectors of lattice $\mu$-th direction. For $x^+_\mu\,\equiv \,+\,ae^\mu$, let $b_\mu(x)\,=\,[x,x^+_\mu]$ denote the positively oriented lattice bond (or, simply, bond) with initial point $x$ and terminal point $x _+^\mu\equiv x + ae^\mu \in\Lambda$. We denote both the collection and the number of the lattice bonds by $\Lambda_b$ in the text (there should be no confusion!). Sometimes, we may refer to the bonds in the time direction $x^0$ as the {\em vertical} bonds. The spatial bonds are said to be {\em horizontal}. 

We define the variables of the model. To each bond $b\in\Lambda$, we associate a unitary matrix variable $U_b\in\mathcal G$. In addition, we define the total gauge group $\mathcal G_t$ which is a product over the gauge groups ${\mathrm SU}(N)$ associated with each lattice site $x$ given by
\bequ\lb{tgg}
\mathcal G_t\,\equiv\,\prod_{x\in\Lambda}\, [{\mathrm SU}(N)]_x\,.
\eequ
The total gauge group $\mathcal G_t$ acts on the bond variables (see Eq. (\ref{gt}) below). When there is no confusion, we will keep using ${\mathrm SU}(N)$ to refer to $\mathcal G_t$.

Let $d\mu_b(U_b)$ denote the normalized Haar measure associated with the gauge group $\mathcal G$ for the bond $b$, then measure for $\mathcal U$ is product measure $d\mu(\mathcal U)\,=\,\prod_{b\in\Lambda}\,d\mu(U_b)$, where $\mathcal U$ denotes the collection of bond variables $U_b$, $b\in\Lambda$.

Here, it is important to call the reader´s attention to the fact that the case of the gauge group ${\mathrm SU}(2)$ is very special since it involves only real characters whilst we have complex characters for ${\mathrm SU}(N\not= 2)$. 

Another important ingredient in the definition of a lattice YM model is a plaquette   $p\subset\Lambda$. Plaquettes are minimal square circuits in $\Lambda$. Precisely, the plaquette $p\equiv p_{\mu\nu}(x)$, in the $\mu\nu$-plane, with $\mu<\nu$, has vertices on the lattice sites $x$, $x^\mu_+\,\equiv\,x + ae^\mu$, $x^\mu_+\,+\,ae^\nu$, $x^\nu_+\,\equiv\,x + ae^\nu$ of $\Lambda$. For $\dagger$ denoting the adjoint and $p\,=\,p_{\mu\nu}(x)$, we define the $\mathcal G$-valued plaquette variable
\bequ
\lb{Upp}
U_p\,\equiv\,U_{b_1}\,U_{b_2}\,U^\dagger_{b_3}\,U^\dagger_{b_4}\,=\,U_{b_1}\,U_{b_2}\,\left[U_{b_4}\,U_{b_3}\right]^\dagger\,\equiv\,e^{iX_p}\,.
\eequ

  We define the $d=4$ Euclidean spacetime dimensional YM model partition function by
\bequ
\lb{partitionB}Z_{\Lambda,a}\,\equiv\,\dis\int \;\exp\left[-\beta\,A_{\Lambda,a}(\mathcal U)\right]\:d\mu(\mathcal U)\,\equiv\,\dis\int \;\exp\left[-\beta\,\dis\sum_{p\in\Lambda}\,A_p(U_p)\right]\:d\mu(\mathcal U)\,.
\eequ
(For spacetime dimension $d\not= 4$, the prefactor $\beta\,=\,g^{-2}$  is replaced by $a^{d-4}\,g^{-2}$.) Here, the  single Wilson plaquette action $A_p(U_p)$ is given by  
\bequ
\lb{WA}A_p(U_p)\,\equiv\,2\,\Re\,\tr\,(1-U_p)\,=\,\|U_p\,-\,1\|^2_{H-S}\,=\, \tr \left( 2\,-\,U_p\,-\,U_p^\dagger\right)\,=\,\tr 2\,-\,\left[ \tr U_p\,+\,\left(\tr U_p\right)^*\right]\,\geq\,0\,.
\eequ 
The symbol $\|\,\mathcal O\,\|_{H-S}\,=\,\left[\tr\,\left(\mathcal O^\dagger\mathcal O\right)\right]^{1/2}$ denotes the Hilbert-Schmidt norm, where $Tr$ is the trace over the gauge group elements, the star $*$ stands for complex conjugation and we recall that $\tr U_p^\dagger=(\tr U_p)^*$.   Letting $M_1,\,M_2$ be matrices in ${\mathrm SU}(N)$. Then $(M_1,M_2)\,\equiv\,Tr(M_1^\dagger M_2)$ is a sesquilinear inner product.

The first equality in Eq. (\ref{WA}) is easily proved by direct computation. 
The last equality for $A_p(U_p)$ uses the fact $U_p$ is a unitary variable, as follows from Eq. (\ref{Upp}). We emphasize that all the above definitions are independent of the gauge variable parametrization.  An specially important parametrization is the {\em physical parametrization} \bequ\lb{physpar}U_b\,=\,\exp[igaA^u_b]\,,\eequ
with $A_b^u$ in the Lie algebra of $\mathcal G$. The gauge field $A^u_b$, with the upper index $u$, is the {\em unscaled} or physical field and has the representation $A^u_b=\sum_{1\leq \alpha\leq \delta_N} A_{b}^{u,\alpha}\, \theta_\alpha$, where the Lie algebra generators $\theta_\alpha$, $\alpha=1,\ldots,\delta_N=(N^2-1)$, are the self-adjoint and traceless matrices normalized by $\tr \theta_\alpha\theta_\gamma\,=\,\delta_{\alpha\gamma}$,  with a Kronecker delta. We refer to $A_{b}^{u,\alpha}$ as the physical color or gauge components of $A^u_b$.

   The action of Eq. (\ref{WA}) is invariant under local transformations of the total gauge group $\mathcal G_t$ of Eq. (\ref{tgg}). Namely, for $x\in Z_t^\infty$ and $h(x)\in{\mathrm SU}(N)_x$,  the bond variable for the bond $b_\mu(x)\,=\,[x,x^+_\mu\equiv x+e^\mu]$ transforms as
\bequ\lb{gt} \barr{c}
U(g_{x+e^\mu,x})
\,\mapsto\,h(x+e^\mu)\,U(g_{x+e^\mu,x})\,h^{-1}(x) \,,
\earr\eequ  
and the adjoint transforms as the adjoint of Eq. (\ref{gt}). Here, the group elements $h(x)$ may be taken to be different for different sites $x$.

It is important to remark that, formally, when the lattice spacing $a\searrow 0$, in Ref. \cite{Gat}, using the Baker-Campbell-Hausdorff formula (see Refs. \cite{GJ,Far,Moore,Sternberg}), it is shown that, for small lattice spacing $a$, the Wilson plaquette action $\left\{\left(a^{d-4}/g^2\right)\,\sum_{p\in\Lambda}\,A_p(U_p)\right\}$ is the Riemann sum approximation to the usual classical smooth field continuum YM action
\bequ
\lb{TrF2}
{\mathcal A}\,=\,\sum_{\{\mu<\nu\}}\,\dis\int_{[-La,La]^d}\;{\mathrm Tr}[F_{\mu\nu}(x)]^2\,d^dx\,\simeq\,\sum_{\{\mu<\nu\}}\:\;\sum_{x\in\Lambda}\;a^d\, {\mathrm Tr}\left\{F_{\mu\nu}^a\,\equiv\,\partial^a_\mu A_\nu(x)-\partial^a_\nu A_\mu(x)+ig[A_\mu(x),A_\nu(x)]\right\}^2\,.
\eequ
Here, $L$ is the number of sites on a side of the lattice $\Lambda$ and we have the finite difference derivatives   $$\partial^a_\mu A_\nu(x)\,=\,a^{-1}\;[A_\nu(x+ae^\mu)\,-\,A_\nu(x)]\,,$$ and
used the notation $ \{\mu<\nu\}\,\equiv\{\mu,\nu=0,...,(d-1)\,/\,\mu<\nu\}$. (Below, we use the symbol $\partial_\mu$ to denote the usual continuum spacetime partial derivatives.) The approximate action of Eq. (\ref{TrF2}) has local cubic and quartic interaction terms in the gluon fields $A_\mu(x)$. However, these terms are not present in the non-naive, exact form that we derive below, in Appendix B. To our surprise, in Ref. \cite{ROMP}, the single plaquette Wilson action turns out to satisfy the global upper quadratic bound
\bequ
\lb{lower1}
A_p\,=\,\|U_p-1\|^2_{H-S}\,\equiv\,|2\,\Re\,\tr (U_p-1)|\,\leq\,C^2\,a^2 \,g^2\,\sum_{1\leq j\leq 4}\, \sum_{c_j=1}^{\delta_N}\,|A^{u,c_j}_{b_j}|^2\quad\:\:,\:\:\quad C=2\sqrt N\,,
\eequ
where $U_{b_j}\,=\,e^{iagA^u_{b_j}}$, $j=1,2,3,4$, are the four bond gauge variables of the plaquette $p$ (see Lemma 1 below).

  In the physical parametrization of Eq. (\ref{physpar}), the plaquette energy $A_p/g^2$ is regular in $g^2$, for all $g^2\geq 0$, and has the quadratic growth bound in the fields.

This quadratic bound does not hold e.g. when we analyze the YM model in the continuum spacetime. This is also the case of the stochastic approach   used in Ref. \cite{Hairer}.

In Appendix B, using the Baker-Campbell-Hausdorff formula, we show the Wilson action is entire jointly analytic in the gluon fields. Explicitly for the gauge group ${\mathrm SU}(2)$, we have an expansion containing terms with up to four gauge fields. In dimension $d=4$, there are no local terms in the expansion. From the renormalization group viewpoint, with canonical scaling, and considering spacetime dimension $d=4$, the terms in Eqs. (\ref{K2}) and (\ref{endK}) are all marginal, except for the $\left[\vec L_{\mu\nu}\cdot\left(\vec R_{\mu\nu}\,+\,\vec T_{\mu\nu}\right)\right]$ and $|\vec L_{\mu\nu}|^4$ which   is irrelevant. (Here, $\vec L_{\mu\nu}$ denotes the strength field antisymmetric second order tensor in the abelian case given by the r.h.s. of Eq. (\ref{TrF2}), setting $g=0$.) There is nothing like a usual positive and relevant local quadratic mass term.

In the scaled gluon field parametrization $U_b\,=\,e^{igA_b}$ (see Ref. \cite{ROMP}), with the {\em scaled} field   \bequ\lb{scaled}A_b\,=\,a^{(d-2)/2}\,A^u_b\,,\eequ
(with no upper index), the quadratic term in the Wilson plaquette action  is (recall $\mu,\nu\,=\,0,1,2,3$ denote the lattice spacetime directions)
$$  
\sum_{x\in\Lambda}\;\sum_{\mu,\,\nu}\; |a\,\partial^a_\mu \vec A_\nu(x)\,-\,a\,\partial^a_\nu \vec A_\mu(x)|^2\,,
$$
with a vector notation in the Lie algebra of $\mathcal G={\mathrm SU}(N)$. Of course, this gives rise to massless gluons.

However, if we take into account the action arising from the exponent of the exponentiated density of the gauge group Haar measure. there is a positive local mass term and the total quadratic action with respect to the gluon field Lebesgue measure is now
$$  
\sum_{x\in\Lambda}\;\sum_{\mu,\,\nu}\; |a\,\partial^a_\mu \vec A_\nu(x)\,-\,a\,\partial^a_\nu \vec A_\mu(x)|^2\,+\,\sum_{x\in\Lambda}\;\sum_{\mu}\;\dfrac{g^2}3\,|\vec A_\mu(x) |^2\,.
$$

A local positive mass term also arises for ${\mathrm SU}(N)$ and the coefficient is proportional to $g^2$ and the group quadratic Casimir operator eigenvalue. This is shown in Appendix C.

In the YM model,   concerning the $a$ parameter dependence, there is more regularity with the use of scaled fields in the YM model. We point out that, for scaled fields (see Ref. \cite{ROMP}), the model obeys thermodynamic and ultraviolet (TUV) stability   bounds. The scaled plaquette field correlations are bounded, in the thermodynamic and the continuum limit $a\searrow 0$ (at least in the subsequential sense), for any $0\,<\,g^2\,\leq\,g_0^2 \,<\,\infty$.

Now, for a function $F(U)$, of a collection of bond variables and their inverses, we have the normalized statistical mechanical averages
\bequ\lb{corr}
\langle\,F(U)\,\rangle_{\Lambda,a}\,=\,\dfrac1{Z_{\Lambda,a}}\;\dis\int \,F(U)\,\exp\left[-\beta\,A(\mathcal U)\right]\:d\mu(\mathcal U)\,,
\eequ
which are used to define the YM model correlations on the lattice $\Lambda$.

In \cite{Sei,Si}, using polymer expansions, the thermodynamic limit $\Lambda\nearrow Z_t^\infty\,\equiv\,[(\mathbb Z+\frac12)\times\mathbb Z^3]$ of correlations is proven to exist. The thermodynamic limit of correlations exists and truncated correlations have exponential tree decay. The limiting correlations are lattice translational invariant. Furthermore, the correlation functions extend to analytic functions in the global coupling parameter   $\beta=1/(g^2)\ll 1$.

Also, for $\beta$ small, in Ref. \cite{Schor1,Schor2,Schor3}, the use of a plaquette field and the corresponding truncated two-plaquette field correlation led to the determination one single and isolated one-glueball state in the underlying quantum mechanical Hilbert space $\mathcal H$ of the model, in the case of the ${\mathrm SU}(2)$ gauge group for three Euclidean spacetime dimensions and with fixed lattice spacing $a$.

The determination of the low-lying energy-momentum is done based on the use of a Feynman-Kac type formula on the lattice, relating inner products of vectors (physical states) in $\mathcal H$ and statistical mechanical correlations of the random lattice bond gauge variables (see Eq. (\ref{FeyKa}) below). From spectral representations for the commuting and self-adjoint energy and momentum operators, the lowest-lying spectral point is related to singularities of the lattice Fourier transform of the truncated two-plaquette field correlation. (In direct space, the singularity of the Fourier transform is related to the spacetime distance exponential decay rate of the correlation, according to the Payley-Wiener theorem \cite{Stein}.)

For the $p$ plaquette variable $U_p$ defined in Eq. (\ref{Upp}), the gauge-invariant plaquette field associated with $p$ is given by 
\bequ\lb{plaqfield}\barr{lll}
\tr {\mathcal F}_{p}(x)&=&\dfrac 1 {a^2g}\,\Im\tr(U_p-1)\,=\,\dfrac {i}{2a^2g}\,\tr\,[U^\dagger_p\,-\,U_p]\,=\,
\dfrac1{a^2g}\,\tr (\sin X_p)\,,\earr
\eequ
where in the last equality we have set $U_p\,=\,\exp\{ iX_p \}$. Here, the point $x$ for the plaquette $p$ is taken to be lowest left corner site of $p$, as in Eq. (\ref{Upp}). In the formal continuum limit $a\searrow 0$, using the physical parametrization given in Eq. (\ref{physpar}), the plaquette field gives the usual antisymmetric, second order classical field strength tensor $F_{\mu\nu}(x)=\partial_\mu A_\nu(x)-\partial_\nu A_\mu(x)-ig[ A_\mu(x),A_\nu(x)]$. The trace of $F_{\mu\nu}(x)$ is of cubic order in the $A$´s. 

We also define the gauge-invariant plaquette field
\bequ\lb{F+}{\mathcal F}_{p+}(x)\,=\,\dfrac 1 {(a^2g)^2}\,\Re\tr(U_p-1)\,=\,\dfrac {1}{2(a^2g)^2}\,\tr\,[U^\dagger_p\,+\,U_p]\,.\eequ
For small lattice spacing $a$, the plaquette field ${\mathcal F}_{p+}(x)$ behaves as $$
{\mathcal F}_{p+}(x)\,\simeq\,\dfrac 1 {(a^2g)^2}\,\tr 1\,-\,\tr\,[\partial^a_\mu A_\nu(x)-\partial^a_\nu A_\mu(x)]^2\,+\,\mathcal O(A^3)\,,$$
and, up to a constant, ${\mathcal F}_{p+}(x)$ is minus the Wilson action of Eq. (\ref{partitionB}). The constant disappears when computing the truncated two-plaquette field correlation.

Setting $\tr {\mathcal F}_{p}(x)\,\equiv \,\tr {\mathcal F}_{p-}(x)$ in Eq. (\ref{plaqfield}), numerical simulation results show the existence of two single glueball states and that the mass of $ {\mathcal F}_{p+}(x)$ is smaller than the mass associated with ${\mathcal F}_{p-}(x)$ correlator (see Refs. \cite{Ochs,Munster,Teper,Seo}). For small $\beta$, we prove that the two masses are asymptotically equal to $[(-4\ln\beta)/a]$. Also, the $\mathcal F_\pm$ generate the entire low-lying energy-momentum spectrum of the model. No other lattice loops are needed, besides the plaquette.

The truncated two-plaquette field correlation is then given by
\bequ\lb{2pt}
 G^T_{\Lambda,a}(x_i,x_j)\,\equiv\,\left\langle \tr\,{\mathcal F}_{p_i}(x_i)\;\tr\,{\mathcal F}_{p_j}(x_j)\right\rangle^T_{\Lambda,a}\,\equiv\,\left\langle \tr\,{\mathcal F}_{p_i}(x_i)\;\tr\,{\mathcal F}_{p_j}(x_j)\right\rangle_{\Lambda,a}\,-\,\left\langle \tr\,{\mathcal F}_{p_i}(x_i)\right\rangle_{\Lambda,a}\;\left\langle\tr\,{\mathcal F}_{p_j}(x_j)\right\rangle_{\Lambda,a}\,.
\eequ
\begin{rema}\lb{rema1}
	For the case of the abelian gauge group $\mathrm U(1)$, 
$$
{\mathcal F}_{p+}\,=\,\dfrac 1 {(a^2g)^2}\,[U^*_p\,+\,U_p]\quad,\quad{\mathcal F}_{p-}\,=\,\dfrac {i} {2a^2g}\,[U^*_p\,-\,U_p]\,,
$$
with small $a$ behavior
$$
{\mathcal F}_{p+}(x)\,=\,\dfrac 1 {(a^2g)^2}\,-\,[\partial^a_\mu A_\nu(x)-\partial^a_\nu A_\mu(x)]^2\,+\,\mathcal O(A^4)\quad,\quad {\mathcal F}_{p-}(x)\,=\,\partial^a_\mu A_\nu(x)-\partial^a_\nu A_\mu(x)\,+\,\mathcal O(A^3).
$$
\end{rema}

From the Feynman-Kac formula, we see that the analysis of symmetries of bond and plaquette field variable correlations are related to the physical particle content of the energy-momentum spectrum in the quantum mechanical Hilbert space $\mathcal H$. This is the reason explaining why, in section II, we define charge conjugation operators acting both on the level of functions of random gauge variables (statistical mechanical averages), as well as an operator in $\mathcal H$. 
Subsequently, we make a detailed analysis of the charge conjugation consequences in the structure of the low-lying particle spectrum in the Hilbert space $\mathcal H$.
We show that, due to the fact that the characters for the groups ${\mathrm SU}(N\not=2)$ are complex, in contrast to the case of ${\mathrm SU}(2)$ treated in  \cite{Schor1,Schor2,Schor3}, the Hilbert space $\mathcal H$ splits into two orthogonal components $\mathcal H=\mathcal H_+\,\oplus\,\mathcal H_-$, one corresponding to the subspace $\mathcal H_+$ of $+1$ eigenvalue for charge conjugation and the other component with eigenvalue $-1$. In particular, the plaquette fields $\mathcal F_{p\pm}(x)$ have eigenvalues $\pm 1$. This is one of the main results of this paper! Its consequence, regarding the mass gap one-million dollar problem, is that there are two single-glueball states, eventually with different masses, in the low-lying energy-momentum spectrum whenever the gauge group is taken to be ${\mathrm SU}(N\not=2)$. Our analysis extends to the abelian (electromagnetic) case of the group ${\mathrm U}(1)$.

  As mentioned before, in \cite{ROMP}, in the YM model in Euclidean spacetime dimensions $d=3,4$, using scaled fields, we obtain more regularity in the lattice spacing $a$. The scaling is noncanonical, depends on the lattice spacing $a$, and can be viewed as a (partial) renormalization defined as a field strength renormalization. For a finite $g_0>0$ and for the gauge coupling $g\in(0,g_0]$ (so that we are {\em not} restricted to $\beta$ small!), we showed diverse finiteness properties. Namely, the boundedness of a normalized free-energy, the plaquette-field generating functions, which leads also to the boundedness of correlations themselves. Surprisingly, correlations are bounded even at coincident points! These finiteness properties persist in the thermodynamic limit and hold uniformly in the lattice spacing $a$. Hence, boundedness is also inherited in the continuum limit $a\searrow 0$. The boundedness properties are valid for the whole class of scaled models defined with the Wilson plaquette action
 with $g\in(0,g_0]$, as is the case of the model considered here.
 
Using the existence of the thermodynamic limit for $0<\beta\ll 1$, below we drop the lower index $\Lambda$ in correlations and assume we have taken the thermodynamic limit. Also, the lattice spacing $a\in(0,1]$ is fixed and the index $a$ will be omitted.

We note that, according to the cofactor method by Creutz (see Refs. \cite{Creu,Creu2}), whenever $g_b^{-1}$ occurs, it can be replaced by combination, with real coefficient, of $(N-1)$ $g_b$´s and the only surviving integrals with the Haar measure $d\mu(U)$ are monomials on the $F$´s with a multiple of $N$ variables $g_b$.

It is also useful to recall some general results given in the Lemma below, which is proved in Ref. \cite{ROMP}. 
\begin{lemma}
	\lb{lemma1}
	\begin{enumerate}
		\item Let $X$ be a self-adjoint, traceless $N\times N$ matrix. Define $\exp(iX)$ by the Taylor series expansion of the exponential.  Then $\exp(iX)\in{\mathrm SU}(N)$.
		\item Given $U\in{\mathrm SU}(N)$, by the spectral theorem, there exists a unitary $V$ such that $V^{-1}UV\,=\, \mathrm{diag}(e^{i\lambda_1}, \ldots,e^{i\lambda_N})$, $\lambda_j\in(-\pi,\pi]$, $j=1,\ldots,(N-1)$ and $\lambda_N\,=\,-\sum_{1\leq j\leq(N-1)}\,\lambda_j$. The $\lambda_j$ are the angular eigenvalues of $U$.  Define $X=V\mathrm{diag}(\lambda_1,\ldots,\lambda_N)V^{-1}$. Then, $X$ is self-adjoint and traceless. Furthermore,  $U= \exp(iX)$, and the exponential map is onto (see \cite{Bump}).
		\item Let $\theta_\alpha$, $\alpha\,=\,1,\ldots, \delta_N\equiv (N^2 -1)$, be the traceless and self-adjoint ${\mathrm SU}(N)$ Lie algebra generators, normalized according to $\tr \theta_\alpha\,\theta_\gamma\,=\,\delta_{\alpha\gamma}$. 
		Let $X$ be self-adjoint and traceless. Thus, $X$ has the representation $X=\sum_{1\leq \alpha\leq \delta_N} \;x_\alpha\theta_\alpha\,=\,\vec x\cdot\,\vec\theta$, with $x_\alpha\,=\,\tr X\theta_\alpha$, for $x_\alpha$ real.
		\item For $U$ and $X$ related as in item $2$, we have the important inequality:
		$$\barr{lll}
		\| X\|^2_{H-S}&=&\mathrm{Tr}\left(X^\dagger X\right)\,=\,\sum_{1\leq\alpha\leq N^2-1}\,|x_\alpha|^2=|x|^2\,=\,\sum_{1\leq  j\leq N}\,\lambda_j^2
		\vspace{2mm}\\
		&=& \left[ \sum_{j=1}^{N-1} \lambda_j\right]^2\,+\,\left[ \sum_{j=1}^{N-1} \lambda_j^2\right]\,\leq\,(N-1)^2\,\pi^2\,+\,(N-1)\pi^2\,\leq\,N(N-1)\,\pi^2\,.
		\earr
		$$
		Thus, the exponential map is onto, for $|x|\leq [N(N-1)]^{1/2}\;\pi$.
		\item In particular, if $X_b\,=\,agA_b^u$, the equality in item 4 takes the form 
		\bequ
		\lb{item5}\| X\|^2_{H-S}\,=\,\mathrm{Tr}\left(X^\dagger X\right)\,=\,\sum_{\alpha=1}^{\delta_N}\,|x_\alpha|^2\,=\,
		\sum_{j=1}^{N}\; \lambda_j^2\,=\, a^2g^2\;\sum_{c=1}^{\delta_N}\, \left|A_b^{u,c}\right|^2\,.
		\eequ
	\end{enumerate}
\end{lemma}

For $p=p_{\mu\nu}(x)$, adopting the physical parametrization for the $U_b$´s in the plaquette variable $U_p$ [see Eq. (\ref{Upp})], and taking $U_p= e^{iX_p}$, the positive plaquette action $A_p$ for the plaquette $p$ defined below Eq. (\ref{partitionB}) can be written as
\bequ
\lb{Ap}
A^u_p\,=\,2\tr(1–\cos X_p)\,.
\eequ

This completes the description of the model and we now consider the YM model underlying quantum mechanical Hilbert space. 

As given above , the Wilson action of Eq. (\ref{WA}) is invariant under local transformations of the total gauge group $\mathcal G_t$ of Eq. (\ref{tgg}). Other symmetries of the action of Eq. (\ref{WA}), such as time reversal, charge conjugation, parity, coordinate reflections and spatial rotations,
which can be implemented by unitary (anti-unitary for time
reversal) operators on the physical Hilbert space ${\mathcal H}$, were
treated in Refs. \cite{CMP,8fdw}. As remarked below, these operators can be order preserving (homomorphisms) or order reversing (anti-homomorphisms). Furthermore, in Refs. \cite{CMP,8fdw}, a new {\em time reflection} symmetry, to be distinguished from time reversal, of the correlations for the model with fermions is used to define a new local spin flip symmetry which is shown to be useful for obtaining additional relations between correlations. 

The   Hilbert space ${\mathcal H}$
and the E-M operators $H$ and $P^j$, $j=1,2,3$ are defined as in
Refs. \cite{CMP,8fdw}. We start from gauge invariant correlations,
with support restricted to $u^0=1/2$ and we let $T_0^{x^0}$,
$T_i^{x^i}$, $i=1,2,3$, denote translation of the functions of
the gauge variables by $x^0\geq0$, $\vec
x=(x^1,x^2,x^3)\in\Z^{3}$. For $F$ and $G$ only depending on
coordinates with $u^0=1/2$, we have the Feynman-Kac formula
\bequ\lb{FeyKa}(G,\check T_0^{x^0}\check T_1^{x^1}\check
T_2^{x^2}\check T_3^{x^3}F)_{{\mathcal H}} =\langle [T_0^{x^0}
T^{\vec x}F]\Theta G\rangle\,\,,\eequ where $T^{\vec
	x}=T_1^{x^1}T_2^{x^2} T_3^{x^3}$ and $\Theta$ is an anti-linear
operator which involves time reflection. Following Ref. \cite{Sei},
the action of $\Theta$ on gauge variables is given by  (here, the bond is understood to be connecting the lattice point $u$ to $v$)
\bequ
\lb{Thetag} 
\Theta\,  f(\left\{g_{uv})\right\})\,=\, f^*(\left\{ g_{(tu)(tv)} \right\})\,,
\eequ
where $t(u^0,\vec u)=(-u^0,\vec u)$, for $u,v\in\Lambda$. 
For simplicity, we usually do not distinguish between gauge variables and their associated Hilbert space vectors in our notation. As linear operators in ${\mathcal 	H}$, $\check T_\mu$, $\mu=0,1,2,3$, are mutually commuting; $\check T_0$ is self-adjoint, with $-1\leq \check T_0\leq1$, and $\check
T_{j=1,2,3}$ are unitary. So, we write $\check T_j=e^{iP^j}$ and
$\vec P=(P^1,P^2,P^3)$ is the self-adjoint momentum operator. Its
spectral points are $\vec p\in{\bf T}^3\equiv (-\pi,\pi]^3$. Since
$\check T_0^2\geq0$, the energy operator $H\geq0$ can be defined by
$\check T_0^2=e^{-2H}$. We call a point in the E-M spectrum
associated with spatial momentum $\vec p=\vec 0$ a mass and, to be
used below, we let ${\mathcal E}(\la^0,\vec \la)$ be the product of the
spectral families of $\check T_0$, $P^1$, $P^2$ and $P^3$. By the
spectral theorem (see Ref. \cite{RS1}), we have
$$
\check T_0=\int_{-1}^1\lambda^0 dE_0(\lambda^0) \quad,\quad \check
T_{j=1,2,3}=\int_{-\pi}^\pi\,e^{i\lambda^j} dF_j(\lambda^j)\,,
$$
so that ${\mathcal E}(\la^0,\vec \la)=E_0(\lambda^0)\prod_1^3\,
F_j(\lambda^j)$. The positivity condition $\langle F\Theta
F\rangle\geq 0$ is established in Ref. \cite{Sei}, but there may be
nonzero $F$'s such that $\langle F\Theta F\rangle= 0$. If the
collection of such $F$'s is denoted by ${\mathcal N}$, a pre-Hilbert
space ${\mathcal H}'$ can be constructed from the inner product $\langle
G\Theta F\rangle$ and the physical Hilbert space ${\mathcal H}$ is the
completion of the quotient space ${\mathcal H}'/{\mathcal N}$.
  
Here, of special interest is the action of {\em  charge conjugation} on gauge fields, which is a symmetry of the Wilson action of Eq. (\ref{WA}).
Generically speaking, symmetry operations are defined on
the gauge algebra. If $b$ is a lattice bond connecting the lattice point $x$ to $y$, a symmetry operation $\mathbb Y$ acts on functions
$f(g_b)$ of the gauge group ${\rm SU}(N)$ by \bequ \lb{sym2}
{\mathbb Y}\,f(g_{b})=({\mathbb Y}f)(g^{\sharp}_{{\mathcal Y}x{\mathcal
		Y}y})\,,\eequ where ${\mathcal Y}$ is a transformation acting on the
coordinates and the operation $^{\sharp}$ acts on the matrix elements of $g$ in a way which depends on the explicit case we consider.

A symmetry of the model is defined to be a symmetry operation
which leaves invariant the action of Eq. (\ref{WA}) and
which has the following property involving the normalized
expectations of gauge invariant functions $F$ of the bond variables $\{U_b\}$:
\bequ \lb{diese} \langle{\mathbb Y}F\rangle=\langle
F\rangle^{\sharp}\,,\eequ where $\langle F\rangle^{\sharp}$ means
either $\langle F\rangle$ or $\langle F\rangle^*$.

Furthermore, the symmetry can be implemented on the quantum
mechanical Hilbert space ${\mathcal H}$ by a unitary or anti-unitary
operator stabilizing the null space ${\mathcal N}$ i.e. such that if
$F\in{\mathcal N}$ then ${\mathbb Y} F\in{\mathcal N}$. The symmetry can be order preserving (homomorphism) such as
\bequ\lb{OP}  {\mathbb
	Y}(M_1M_2)={\mathbb Y}(M_1){\mathbb Y}(M_2)\,, \eequ or
order reversing (anti-homomorphism)
 \bequ\lb{OR} {\mathbb
	Y}(M_1M_2)={\mathbb Y}(M_2){\mathbb Y}(M_1)\,. \eequ

We now give a  discussion of {\em Charge Conjugation}. In \cite{CMP,8fdw}, charge conjugation is constructed as an order reversing operation whereas e.g. the usual parity operation, reversing the sign of spatial vector components, is order preserving.

In subsection II.1, a charge conjugation linear operator $\mathcal C$ acting on the quantum mechanical physical Hilbert space $\mathcal H$ is defined and its properties are established. In subsection II.2, we define a charge conjugation operator $\mathcal C_E$ on the function space of correlations. The connection between these two operators is given by the Feynman-Kac formula (see Refs. \cite{GJ,Sei}).
\subsection{Charge Conjugation Operator $\mathcal C$ in the Physical Hilbert Space $\mathcal H$}
\lb{2.1}
Recalling that we adopt the hypercubic lattice $\Lambda\,\subset a\left[\,Z_t\equiv\left\{\pm 1/2,\pm 3/2,\pm 5/2...\right\}\times \mathbb{Z}^{d-1}\right]$,  we now consider complex matrix functions which depend on the matrix elements of the bond variables $U_b$. We take the support of the bond $b$ to be $u\,=\,(u^0=1/2\,,\,\vec u)$. Below, we define bounded linear operators on a dense set of $\mathcal H$ and then extend the domain to all of $\mathcal H$ by continuity.

The charge conjugation linear operator ${\mathcal C}$ in the Hilbert space $\mathcal H$ and acting on a function $F$ of the sets of bond variables $U_b$ in $\mathcal U$ and is defined by  (see \cite{CMP,8fdw})
\bequ\lb{CC} 
{\mathcal C}\,F(\left\{U_{b}\right\})\,= \,F\left(\left\{U^*_{b}\right\}\right)\,,
\eequ
and we have ${\mathcal C}\,\Theta\,=\,\Theta\,{\mathcal C}$, with $\Theta$ defined in Eq. (\ref{Thetag}). The action of ${\mathcal C}$ on a product of functions $F_1$, $F_2$, ... is order reversing.

The Wilson action $A(\mathcal U)$ of Eq. (\ref{WA}) is invariant under $\mathcal C$ and we now show that the operator $\mathcal C$ is unitary. To do so, it is enough to show that $\mathcal C$ is isometric (not necessarily onto), that is ${\mathcal C}^\dagger{\mathcal C}\,=\,1$ and noting that the relation ${\mathcal C}\,{\mathcal C}\,=\,1$ shows that $\mathcal C$ is onto. Thus, ${\mathcal C}^\dagger\,=\,{\mathcal C}^{-1}$ and ${\mathcal C}\,{\mathcal C}^\dagger\,=\,1$. Furthermore, ${\mathcal C}^{-1}\,=\,{\mathcal C}^\dagger\,=\,{\mathcal C}$ and ${\mathcal C}$ is self-adjoint.

We write $F$ as a finite linear combination of multinomials in the matrix elements of the $g_b$´s. Namely, for $1\leq j\leq r$, we have
\bequ\lb{LCF}
F\,=\, \dis\sum_{\{n_j\}\,,\,\{j_\ell,k_\ell\}}\; c_{b_1,j_1,k_1,\ldots,b_r,j_r,k_r}\; g^{n_1}_{b_1j_1k_1}\dots g^{n_r}_{b_rj_rk_r}\,,
\eequ
where $0\leq n_j<\infty$ and $1\leq i_\ell,j_\ell\leq N$. By polarization, it is enough to show the following equality between inner products in $\mathcal H$
$$
\left( {\mathcal C}F,{\mathcal C}F\right)_{\mathcal H}\,=\,\left( F,F\right)_{\mathcal H}
$$
which, using the Feynman-Kac formula of Eq. (\ref{FeyKa}) gives
\bequ\lb{Ccorrel}
\langle  {\mathcal C}F\,\Theta\,{\mathcal C}F\rangle_{\Lambda,a}\,=\,\langle  F\,\Theta\,F\rangle_{\Lambda,a}\,.
\eequ

Recall that the action $A(\mathcal U)$ is invariant under $\mathcal C$. For $A(\mathcal U)=0$, the r.h.s. of Eq. (\ref{Ccorrel}), using Eq. (\ref{LCF}) is
\bequ\lb{Ccorrel2}\barr{lll}
\left( {\mathcal C}F,{\mathcal C}F\right)_{\mathcal H}&=& \dis\int\,F( g^*_{xy}) \,F^*( g^*_{x(ty)}) \,d\mu(\mathcal U)  \vspace{2mm}\\
&=& \dis\int\, \left[ \dis\sum_{\{n_j\}\,,\,\{j_\ell,k_\ell\}}\; c_{b_1,j_1,k_1,\ldots,b_r,j_r,k_r}\; \left[g^{n_1}_{b_1j_1k_1}\right]^*\dots \left[g^{n_r}_{b_rj_rk_1}\right]^*  \right]\vspace{2mm}\\&&\;\;\;\;\;\times\;\left[ \dis\sum_{\{n_j\}\,,\,\{j_\ell,k_\ell\}}\; c^*_{b_1,j_1,k_1,\ldots,b_r,j_r,k_r}\; g^{n_1}_{\left(tb_1\right)j_1k_1}\dots g^{n_r}_{\left(tb_r\right)j_rk_1}  \right]\;\,d\mu(\mathcal U)\,.

\earr
\eequ
By the Creutz cofactor method \cite{Creu,Creu2}, the integral over each $g_b$ is real, so that in Eq. (\ref{Ccorrel2}) we can replace $g_b^\dagger$
by $g_b$ and $g_{tb}$ by $g_{tb}^*$. Comparing with $\langle  F\,\Theta\,F\rangle_{\Lambda,a}$, we see that Eq. (\ref{Ccorrel}) is satisfied.

As noted earlier, as $\mathcal C$ is both unitary and self-adjoint, it has eigenvalues $\pm 1$. Then, it follows that the operators \bequ\lb{Ps}P_\pm\,=\,\left(1\,\pm\,\mathcal C\right)/2\,.\eequ are mutually orthogonal, orthogonal projectors. That is,
\bequ\lb{project}
(P_\pm)^2\,=\,P_\pm,\quad P_+P_-\,=0,\quad P_-P_+\,=\,0,\quad P^\dagger_\pm\,=\,P_\pm\,.
\eequ

With these results, we see that the Hilbert space $\mathcal H$ admits an orthogonal direct sum decomposition into charge conjugation $+$ and $-$ sectors. Namely, with $\mathcal H_\pm\,=\,P_\pm \mathcal H$, we have,
\bequ\lb{Hdecompo}
\mathcal H\,=\,\mathcal H_+\,\oplus \,\mathcal H_-\,.
\eequ
In particular, the gauge-invariant plaquette fields
\bequ\lb{pfields}
({\mathcal F}_p)_\pm\,=\, \tr U_p
\,\pm\,\tr U_p^\dagger\,=\, \tr U_p
\,\pm\,[\tr U_p]^*\,,
\eequ
satisfy
\bequ\lb{actionp}
\mathcal C \,({\mathcal F}_p)_\pm\,=\,\,=\,\pm\,({\mathcal F}_p)_\pm,\quad P_\pm ({\mathcal F}_p)_\pm\,=\,({\mathcal F}_p)_\pm\,,
\eequ
so that
\bequ\lb{HF}
\left(  ({\mathcal F}_{p_1})_+,({\mathcal F}_{p_2})_-\right)_\mathcal H\,=\,\left(  (P_+{\mathcal F}_{p_1})_+,P_-({\mathcal F}_{p_2})_-\right)_\mathcal H\,=\,0\,.
\eequ

The above argument extends to other correlations to show that 
\bequ\lb{FT}\barr{l}
\left\langle  \prod_{k=1}^{2n+1}\,({\mathcal F}_{p_k})_+\;\prod_{\ell=1}^r \,({\mathcal F}_{p_\ell})_- \right\rangle\,=\,0\qquad,\qquad\left\langle  \prod_{k=1}^{n}\,({\mathcal F}_{p_k})_+\;\prod_{\ell=1}^{2r+1} \,({\mathcal F}_{p_\ell})_- \right\rangle\,=\,0\,.
\earr
\eequ
This is a type of non-abelian Furry´s Theorem in quantum electrodynamics \cite{JR}.

Of course, the decomposition of the Hilbert space $\mathcal H$ has major consequences in the analysis of the particle, energy-momentum spectrum of the YM model and the physically expected YM mass gap property: two glueball particles are expected to occur, at least for small $\beta$, leading to two not necessarily equal mass gaps.
\subsection{Charge Conjugation Operator $\mathcal C_E$ Acting on Functions of the Gauge Variables and Correlations}
\lb{2.2}
Here, to finish our analysis of charge conjugation, we define a charge conjugation operator $\mathcal C_E$ in the space of functions of the sets of $g_b$ variables in $\mathcal U$ of the form of Eq. (\ref{LCF}), but with no restrictions on their support. Explicitly, $\mathcal C_E$ is the order reversing linear operator given by (recalling $*$ denotes complex conjugation)
\bequ\lb{CE}
\mathcal C_E f(\{g_b\})\,=\,f(\{g_b^*\}\})\,,
\eequ
so that  $(\mathcal C_E)^2\,=\,1$.

Using this definition and the order reversing character of $\mathcal C_E$, we have  $\mathcal C_E (\tr U_p)\,=\, \tr U_p^\dagger$ and, since $\mathcal C_E^2=1$, also to $\mathcal C_E(\tr U_p^\dagger)\,=\, \tr U_p$. Using these relations when computing the action of $\mathcal C_E$ on the system action $A_p(U_p)$ of Eq. (\ref{WA}), gives that $\mathcal C_E$ is a symmetry of our model.

We equip the space of functions with the inner product
$$
(G,F)_E\,=\, \langle G^* F\rangle\,,
$$
which, upon completion of the quotient space, becomes a Hilbert space $\mathcal H_E$. The norm of a generic vector $F$ is
$$
\|\,F\,\|_E\,=\, \langle F^* F\rangle^{1/2}\,.
$$

Following the arguments given in the previous subsection, $\mathcal C_E\,:\,\mathcal H_E\rightarrow \mathcal H_E$ is an isometric operator, i.e. 
$\mathcal C_E^\dagger\mathcal C_E\,=\,1$, and the relation $\mathcal C_E^2\,=\,1$ shows it is unitary and $\mathcal C_E^\dagger\,=\,\mathcal C_E^{-1}\,=\,\mathcal C_E$.

As $\mathcal C_E$ is both unitary and self-adjoint, $\mathcal C_E$ has eigenvalues $\pm 1$. Hence, as before, we have the orthogonal direct sum decomposition of $\mathcal H_E$ into subspaces $\mathcal H_{E,\pm}$ where, with projectors $P_{E,\pm}$ defined, as above, by $P_{E,\pm}\,=\, (1\,\pm\,\mathcal C_E)/2$,
$$\mathcal H_{E,\pm}\,=\,P_\pm \mathcal H_{E}\,.$$

The orthogonality property is used to treat the gauge-invariant plaquette correlations $({\mathcal F}_p)_\pm$ of Eq. (\ref{pfields}) independently in determining the spacetime decay property analysis of their truncated correlations. In this case, we have $\langle ({\cal F}_{p_1})_+\;({\cal F}_{p_2})_-  \rangle\,=\,0$ and there are {\em no restrictions} on the supports of the plaquettes $p_1$ and $p_2$. Of course, the Furry´s type theorem of Eq. (\ref{FT}) also holds for correlations with no restriction on the support of the plaquettes $p_1$, $p_2$,etc

It is important to emphasize that the orthogonal decomposition of the Hilbert space $\mathcal H$ is {\em not} present in the case of the gauge group ${\mathrm SU}(2)$. This property occurs due to the fact ${\mathrm SU}(2)$ has only real characters. Hence $\tr U_p$ and $(\tr U_p)^*$ are equal and only one charge conjugation sector is present. This is why Refs. \cite{Schor1,Schor2,Schor3} did not have to deal with this multiple sector question in the lattice ${\mathrm SU}(2)$ YM model with an ultraviolet cutoff. There, for fixed lattice spacing $a=1$ and small $\beta$, the use of the basic plaquette fields $\mathcal F_\pm$ of Eq. (\ref{pfields}) together with the hyperplane expansion method (see Refs. \cite{Sp,Si}) was employed to obtain the large distance exponential decay of the truncated two-plaquette field correlation. Using spectral representations for the self-adjoint energy and momentum operators acting on $\mathcal H$, a rigorous relation between singularities of a lattice Fourier transform of the truncated two-plaquette correlation and its spacetime decay properties is made, according to Payley-Wiener theorem \cite{Stein}. The exponential large distance decay rate corresponds to the zero spatial momentum value of the single glueball dispersion curve. A single mass gap of $-4\ln\beta\,+\,r(\beta)$, for $0<\beta\ll 1$ is obtained and $r(\beta)$ is shown to be analytic.

  Again, the two-plaquette lattice circuit is enough to control the short-distance behavior of the convolution inverse of a certain lattice Fourier transform of the two-plaquette correlation. No longer circuits on the lattice are needed! Using the same type of techniques, together with the Hilbert space decomposition results of the present paper, in Ref. \cite{PMR} and considering the gauge group ${\mathrm SU}(N\not= 2)$, we showed there are two glueball states with asymptotic mass $-4\ln\beta\, +\,r_i(\beta)$, $i=1,2$, with $r_i(\beta)$ analytic, for $\beta\ll 1$. The final goal is to take the continuum limit $a\searrow 0$ and obtain the   same results, showing that a mass gap persists in the continuum limit, as well as to extend the existence of the mass gaps for larger values of $\beta$. But, this is the subject of further investigation.
\section{Elitzur's Theorem Revisited and Applications}\lb{sec3}
The most well-known result of Elitzur´s theorem \cite{Eli} is that the expectation of the non gauge-invariant function $g_b$ is zero. That is, we have $$\langle g_b\rangle\,=\,0\,.$$ 

Here, we give a version of Elitzur´s theorem in the theorem below, which is proved in Appendix A. Some examples of its use are also presented below. Our version for Elitzur´s theorem is as follows.
\begin{thm}
	Let $\{b_k\}$ denote the finite set of distinct bonds of a finite collection of disjoint trees in the lattice $\Lambda\,=\,Z_t$ (or the infinite lattice $Z_t^\infty$). Let $F_j$ be a complex $L^2$ function of a single bond variable of the gauge group and let $\mathcal U$ denote the collection of all the bond variables. Then, we have
	\bequ\lb{Elit}\barr{lll}
	\left\langle \dis\prod_{j}\,F_j(g_j)\right\rangle&\equiv&\dis\dfrac1Z\,\int\,\dis\prod_{j}\,F_j(g_j)\;e^{-A_\Lambda(\mathcal U)}\,d\nu(\mathcal U)\vspace{2mm}\\
	&=&\dis\prod_{j}\,\int\,F_j(g_j)\,d\mu(g_j)\,\equiv\,\dis\prod_{j}\,\langle F_j(g_j)\rangle_0\,.
	\earr\eequ
	The measure $d\nu(\mathcal U)$ is the product of normalized single bond Haar measures on $\mathcal G$ and the null index on the average means the total action $A_\Lambda(\mathcal U)$ of Eq. (\ref{partitionB}) is set to zero.
\end{thm}

We now give some examples of more general applications of this theorem.
\begin{enumerate}
	\item Let $\mathcal G\,=\,\mathrm U(1)$, for any bond $b$, and write $g_b\,=\,e^{iA_b}$ such that $\sin A_b\,=\,[g_b- g^*_b]/(2i)$. Then, by the $A_b\,\longrightarrow\,(-A_b)$ symmetry of the action and the Haar measure, we have
	$$ \langle\, \sin A_b\,\rangle\,=\,0\,,$$
	and, by the factorization formula given in Appendix A [see Eq. (\ref{Eli2})],
	$$
	\langle\, \sin A_{b_1}\,\sin A_{b_2}\,\rangle\,=\,	\langle\, \sin A_{b_1}\,\rangle_0\,\langle\,\sin A_{b_2}\,\rangle_0\,=\,0\quad ,\quad b_1\,\not=\,b_2\,,
	$$
	which is a quite surprising result. The correlation is {\em not} gauge invariant.
	
	Moreover, adopting the lattice Fourier transform convention
	$$
	f(x)\,=\,\dfrac1{2\pi}\,\sum_{n\in\mathbb Z}\,a_ne^{inx}\quad,\quad a_n\,=\,\dis\int_{-\pi}^{\pi}\,f(x)\,e^{-inx}\,dx\,,
	$$
	taking $f(x)\,=\,x$, we can define an approximate $A^\eta_b$ by its Fourier series (recall that, by hypothesis, we are dealing with $L^2$ functions!), i.e.
\bequ\lb{Ab}
	A^\eta_b\,=\,\left\{ \barr{cll}
	-\,\dis\sum_{n=1}^\eta\!\,\dfrac {(-1)^n\,[g_b^n\,-\,( g_b^*)^n]}{in}&,&g_b\not=-1\,,\vspace{2mm}\\
	0&,& g_b=-1\,.
	\earr \right.
 \eequ
 $A^\eta_b$ is a random variable. In the limit $\eta\nearrow\infty$, the above series converges pointwise to the linear function $A_b$, for $-\pi< A_b< \pi$,  $g\not=-1$. This function is periodic and corresponds to a truncation of the principal branch of $[log (g_b)]/i$. The series also converges in $L_2((-\pi,\pi))$. 
 
With the definition of the random variable $A^\eta_b$ in force, Elitzur´s theorem implies the vanishing of the non gauge-invariant average
\bequ\lb{AbAb}
\langle\, A^\eta_{b_1}\,A^\eta_{b_2}\, \rangle\,=\,0\quad ,\quad b_1\,\not=\,b_2\,.
\eequ

In Appendix A, the expectations of the $A^\eta_{b}$´s was considered when proving Elitzur´s Theorem [see Eqs. (\ref{avg1}), (\ref{avg2}), (\ref{avg3}) below].  In Eq. (\ref{AbAb}), we have the product of two random variable and use their joint probability distribution. The result given in Eq. (\ref{AbAb}) holds since we can apply Elitzur´s Theorem.


Using the $L^2$ restriction again, we obtain the limit $$\langle\, A_{b_1}\,A_{b_2}\, \rangle\,\equiv\,\lim_{\eta\nearrow \infty} \,\langle\, A^\eta_{b_1}\,A^\eta_{b_2}\, \rangle\,=\,0\,.$$
This result extends to the Furry type Theorem (see Eq. (\ref{FT}))
\bequ\lb{EFT}\barr{l}
\langle\, \prod_{k=1}^{2n+1}A^\eta_{b_k}\,\prod_{\ell=1}^{r}\,A^\eta_{b_\ell}\, \rangle\,=\,0\,.
\earr
\eequ

From the above, defining the approximate non-gauge invariant electromagnetic antisymmetric strength field tensor in terms of the lattice vector potentials ($em$ here stands for electromagnetic!)
\bequ\lb{F1}
F_{\mu\nu}^{em}(x)\,=\,a\, \partial^a_\mu A_\nu(x)\,-\,a\,\partial^a_\nu A_\mu(x)\,=\,[A_\nu(x_\mu^+)\,-\,A_\nu(x)]\,-\,[A_\mu(x_\nu^+)\,-\,A_\mu(x)]\,,
\eequ
we have the even more surprising result
\bequ\lb{surp}
\langle\, F_{\mu\nu}^{em}(x)\,F_{\rho\sigma}^{em}(y)\,\rangle\,=\,0\,,
\eequ
if the plaquettes $p_{\mu\nu}(x)$ and $p_{\rho\sigma}(y)$ are disjoint. Note that $F_{\mu\nu}^{em}(x)$ here is not to be confused with the gauge invariant correlator $\mathcal F_{p-}(x)$ of Remark \ref{rema1}.

\item Let $\mathcal G\,=\,\mathrm{SU}(2)$.  Using the parametrization $g\,=\,x_0\,1\,+\,\vec x \cdot \vec{\sigma}\,$,  $[(x^0,\vec x)\,\in\,\mathbb R^4$, with $[\,(x^0)^2\,+\,|\vec x|^2\,]\,=\,1$, we have that $\mathrm{SU}(2)$ is diffeomorphic to the sphere $S^3$. With $\vec A\,=\,(A^1,A^2,A^3)\in \mathbb R^3$, the physical gluon representation is given by
\bequ\lb{repsu2}
g\,=\,e^{i\vec A\cdot\vec\sigma}\,=\, \cos |\vec A|\,+\,i\,\vec A\cdot\vec\sigma\;\dis\dfrac{\sin |\vec A|}{|\vec A|}\,.
\eequ
For $|\vec A|\,<\,\pi$, the map $\vec A\longrightarrow g$ is $1-1$ and the image is $\mathrm{SU}(2)$ except for $g=-1$. For $|\vec A|\,=\,\pi$, $g=-1$, the map is not $1-1$. For $|\vec A|\,\leq\,\pi$, the map is onto. The Haar measure is (see Refs. \cite{GJ,Molin})
	\bequ\lb{haarsu2}
	d\mu(g)\,=\, \dis\dfrac1{2\pi^2}\,\dfrac{\sin^2 |\vec A|}{|\vec A|^2}\,d\vec A\,=\,\dis\dfrac1{2\pi^2}\,\, \left\{ \exp\left[ -\,\sum_{k\geq 1}\;\dfrac{\zeta(2k)\,|\vec A|^{2k}}{\pi^{2k}\,k}  \right] \right\}^2\,d\vec A\quad,\quad |\vec A|\,\leq\,\pi\,,
	\eequ
	where $\zeta(z)\,=\,\sum_{n\geq 1}\,n^{-z}$ is the Riemann zeta function.
	
	The above series coefficients are all positive, starting with a positive quadratic mass term. The coefficients are related to the Bernoulli numbers \cite{Moore}.
	The first terms are $[\,|\vec A|^2/6\,+\,|\vec A|^4/180\,]$.
	We obtain an explicit closed expression for the components of $\vec A$, as a function of $g$. Using Eqs. (\ref{repsu2}) and (\ref{haarsu2}), and since $\tr\sigma^j\,=\,0$, for $j=1,2,3$, we have 
	\bequ\lb{trsu2}
	\dfrac12\,\tr\,(1\,-\,g)\,=\,\dfrac12\,\tr\left(1-\,e^{i\vec A\cdot\vec\sigma}\right)\,=\,\left(1- \cos |\vec A|\right)\,=\,2\,\sin^2[|\vec A|/2]\,.
	\eequ
	Hence, 
	\bequ\lb{gluonsu2}
	|\vec A|\,=\,2\,\arcsin \left[\,\dfrac{\tr(1-g)}4\, \right]^{1/2}\,,
	\eequ
	where $\arcsin u\,=\, u\,+\,u^3/6\,+\, 3u^5/40 \,+\, 15u^7/ 336\,+\,\ldots$, for $0\leq u^2\leq 1$.
	
	Also, for $g\not=-1$ and $j=1,2,3$, putting \bequ\lb{modA}\tr[\sigma^j\,g/(2i)]\,=\,\dfrac{\sin |\vec A|}{|\vec A|} A^j\,\equiv\,f^j(g)\,,\eequ we have that $\sum_{j=1,2,3}\,[f^j(g)]^2\,=\,\sin^2 |\vec A|$.
	
	With this, by defining
	\bequ\lb{Aj}
	A^j\,=\,\left\{ \barr{cll} 2\,f^j(g)\,\dfrac{\arcsin\left[\tr(1-g)/4 \right]^{1/2}}{\left[\sum_{j=1,2,3}\,[f^j(g)]^2\right]^{1/2}}&,&g\not=\pm1,\vspace{2mm}\\0&,&   g=\pm1\,,
	\earr \right.
	\eequ
	from the evenness of the Haar measure, we obtain
	\bequ\lb{avA}
	\langle\,A^j_\mu(x)\,\rangle\,=\,0\quad,\quad \mu=0,1,\ldots,d\,\,,
	\eequ
	and Elitzur´s theorem gives, for $b\not= \hat b$ or $(\mu\,x)\not=(\nu\,y)$,
	\bequ\lb{avAA}
	\langle\,A^j_\mu(x)\,A^k_\nu(y)\,\rangle\,=\,0\,.
	\eequ
\begin{rema}
	$A^{1,3}$ has eigenvalue $-1$ under charge conjugation while $A^2$ has eigenvalue $+1$. These results emerge from Eqs. (\ref{modA}) and (\ref{Aj}).
\end{rema}

	\item Take the gauge group $\mathcal G=\mathrm{SU}(N)$, for $N\geq 3$. In this case, we do not know of any explicit closed form expression for $A^j$, $j\,=\,1,\ldots,(N^2-1)$. (The series coefficients are given in formula (\ref{pull}) below.) Nor do we have a minimal domain of the $A^j$ that covers $\mathrm{SU}(N)$. However, we can show that  $|A^j|$ is bounded.
	
	For $\mathcal G=\mathrm{SU}(N\geq 3)$, we express $A_b^c$ by an approximate field in terms of $g_b$, which we call $B^c_b$. We give another Furry typr Theorem for this case. Recalling that $g_b\,=\,e^{\vec \theta\cdot\vec\tau}$, with a normalized LIe algebra basis $\vec\tau$ satisfying $\tr \tau^j\,=\,0$ and $\tr\,\tau^j\tau^k\,=\,\delta^{jk}$, and $A_b$ has real color components $A_b^c$. Now,
$$
	g_b\,=\,\cos \vec\tau\cdot\vec A_b\,+\,i \sin\vec\tau\cdot\vec A_b
	$$
	so that
	$$
	\dfrac1{2i}\,\left(g_b\,-\,g_b^\dagger\right)\,=\, \sin\vec\tau\cdot\vec A_b\,=\,\vec\tau\cdot\vec A_b\,+\,\left[ \sin\vec\tau\cdot\vec A_b \,-\,\vec\tau\cdot\vec A_b\,\,-\,\dis\sum_{k=0}^\infty\,\dfrac{(-1)^k}{(2k+1)!}\,(\vec\tau\cdot\vec A_b)^{2k+1}\right]\,.
	$$
	
	Now, consider the approximate $B_b^c$ defined by
	$$\barr{lll}
	B_b^c&=&\tr 	\dfrac1{2i}\,\left(g_b\,-\,g_b^\dagger\right)\,\tau^c\,=\,\tr \sin\vec\tau\cdot\vec A_b\tau^c\vspace{2mm}\\&=&
	\tr(\tau^c\tau^a) A_b^a\,+\tr\left\{\tau^c\,-\,\dis\sum_{k=0}^\infty\,\dfrac{(-1)^k}{(2k+1)!}\,(\vec\tau\cdot\vec A_b)^{2k+1}\right\}\vspace{2mm}\\&=&A^c_b\,+\,\mathcal O(A_b^3)\,.
	\earr
	$$
	
	With this, applying Peter-Weyl Theorem, we obtain
	$$
	\dis\int\,B^c_b\,d\mu(g_b)\,=\,\dfrac1{2i}\,\tr \tau^c\dis\int\,\left(g_b\,-\,g_b^\dagger\right)\,d\mu(g_b)\,=\,0\,.
	$$
	
	Hence, by Elitzur´s Theorem, we have the Furry type Theorem
	$$
	\left\langle\,\prod_{\ell=1}^n B_{b_n}^{c_n}    \right\rangle\,=\,0\qquad,\qquad {\mathrm for}\:\; b_n\:\; {\mathrm all\:\; disjoint\,.}
	$$
	
	By charge conjugation, we also have
	$$\lb{EFT2}\barr{l}
	\left\langle\, \prod_{k=1}^{2n+1}{\mathcal F}_{p_k+}\,\prod_{\ell=1}^{r}\,{\mathcal F}_{p_\ell-}\, \right\rangle\,=\,0\qquad ,\qquad
	\left\langle\, \prod_{k=1}^{n}{\mathcal F_{b_k+}}\,\prod_{\ell=1}^{2r+1}\,{\mathcal F}_{b_\ell-}\, \right\rangle\,=\,0\,,
	\earr
    $$
    with no restriction on the supports of the plaquettes.	
	\item An important composed field in YM theory on the lattice is the gauge-invariant plaquette field. Let us take the plaquette $p\,=\,p_{\mu\nu}(x)$, $\mu<\nu$, in the $\mu\nu$ coordinate plane. Then, upon setting the plaquette variable $U_p$ defined in Eq. (\ref{Upp}) as  $U_p\,=\,\exp\{ iX_p \}$,  the plaquette field corresponding to $p$ is given by, 
    \bequ\lb{plaqfield2}\barr{lll}
    \tr {\mathcal F}^u_{p}(x)&=&\dfrac 1 {a^2g}\,\Im\tr(U_p-1)\,=\,\dfrac {i}{2a^2g}\,[U^\dagger_p\,-\,U_p]\,=\,
    \dfrac1{a^2g}\,\tr (\sin X_p)\,.\earr
    \eequ
    
    The reason for the above multiplicative $[1/(a^2g)]$ factor in the model action [see Eq. (\ref{partitionB})] is that, if one uses the physical parametrization for $U_b\,=\,\exp(igaA_b)$ then, for sufficiently small lattice spacing $a$, we obtain that $\tr \mathcal F^u_p\,\simeq\,\tr F_{\mu\nu}^a$, where
    \bequ\lb{fmunua}F_{\mu\nu}^a(x)\,=\,\partial^a_\mu A_\nu(x)-\partial^a_\nu A_\mu(x)+ig[A_\mu(x),A_\nu(x)]\,,\eequ
    which agrees with the expression for the classical field strength antisymmetric tensor in terms of the lattice vector potentials $A_\mu(x)$. Here, the commutator is taken in the Lie algebra of $\mathcal G\,=\,\mathrm{SU}(N)$ and $\partial^a$ stands for the finite difference derivative. 
    
    By Elitzur´s theorem given above, $\langle   {F}^u_{p_1}(x)\, {F}^u_{p_2}(y)\rangle\,=\,0$, for $p_1\not =p_2$.
\end{enumerate}

\section {Concluding Remarks}\lb{sec4}
We consider a four spacetime Euclidean dimensional Wilson lattice Yang-Mills model with gauge group ${\mathrm SU}(N)$   and small gauge coupling $0<\beta=(1/g^2)\ll 1$ multiplying the Wilson plaquette actions $A_p$. The underlying quantum mechanical Hilbert space $\mathcal H$ is constructed according to the Osterwalder-Schrader-Seiler prescription and a Feynman-Kac formula is established. In the Wilson model, to each lattice bond $b$ there is assigned a bond variable $U_b\in{\mathrm SU}(N)$.
The vector gluon fields are parameters in the Lie algebra of ${\mathrm SU}(N)$. 

It is known that, in the gluon field parametrization, the action is bounded (from above!) quadratically in the gluon fields. In the gluon field expansion of the action, there are no local terms. Hence, there is no local mass term. However, the action associated with the exponent of the exponentiated Haar measure density has a local positive mass term which is proportional to the square of the coupling and the ${\mathrm SU}(N)$ quadratic Casimir operator eigenvalue.

For ${\mathrm SU}(2)$, in section \ref{sec3} (see Eq. (\ref{haarsu2})), we give a closed form expression of the Haar measure in terms of Riemann zeta function. It involves a series with positive coefficients which are related to Bernoulli numbers.  The exponential of the measure density gives a mass term.
A deep question is that if this property is manifest for the gauge groups ${\mathrm SU}(N\not= 2)$. Although the appearance of this quadratic mass term does not prove the mass gap property, it indicates a mass is developed. Of course, we must prove the mass subsists the renormalization procedure allowing to take the continuum limit $a\searrow 0$. Using spectral representations for the self-adjoint and commuting energy and momentum operators, the spectral mass gap is associated with the exponential distance decay of the truncated two (gauge-invariant) plaquette field correlations (see Ref. \cite{Schor1,Schor2,Schor3}) via the Payley-Wiener theorem \cite{Stein}.

Below, in Appendix C, using the Cartan-Maurer form, we obtain an  expansion in terms of the gluon fields that is entire, but we have not determined the sign of the non-quadratic terms. This expansion extends previous results valid only for gluon fields near the identity. The question arises on how this property of existence of a mass term is related to the geometry of gauge group. 

We define a charge conjugation operator $\mathcal C$ in the underlying physical quantum mechanical Hilbert space $\mathcal H$ and prove that $\mathcal H$ admits an orthogonal decomposition into sectors with charge conjugation $\pm 1$. In the space of correlations, a charge conjugation operator ${\mathcal C}_E$ is defined and an analogous decomposition holds. Besides, a version of the Elitzur´s theorem is proven and applications are given. It is proven that the expectation averages of two distinct vector potential correlators is zero. 

Of course, the orthogonal decomposition of the underlying Hilbert space $\mathcal H$ has consequences in the analysis of the truncated two-point plaquette field correlation and towards solving the Yang-Mills mass gap problem. For the gauge group ${\mathrm SU}(N\not= 2)$, two glueball states occur, at least for small $\beta$, leading to two possibly different mass gaps.

In closing, we observe that the orthogonal decomposition is trivial in the special case of the gauge group ${\mathrm SU}(2)$. This is true because ${\mathrm SU}(2)$ has only real characters and just the charge conjugation sector with $+1$ eigenvalue is present. This explains why, in the pioneering work by Schor \cite{Schor1,Schor2,Schor3}, working with a fixed lattice spacing, the use of the basic plaquette field $\mathcal F_+$ of Eq. (\ref{pfields}) is enough to determine the existence of glueballs and the mass gap. To do this, the hyperplane expansion method (see Refs. \cite{Sp,Si}) was applied to obtain the exponential decay rate of the truncated two-plaquette field correlation, to determine one glueball state dispersion curve with a corresponding mass gap of $-4\ln\beta\,+\,r(\beta)$, for $0<\beta\ll 1$, with $r(\beta)$ analytic. The splitting of the Hilbert space $\mathcal H$ was {\em not} present. It is important to observe that the two-plaquette lattice circuit is sufficient to control the short-distance behavior of the convolution inverse of a certain lattice Fourier transform of the truncated two-plaquette correlation. No longer circuits on the lattice are needed. Using the same techniques, together with the charge conjugation decomposition results of the Hilbert space $\mathcal H$ proven in the present paper, in Ref. \cite{PMR} we showed there are two one-glueball states with asymptotic mass $-4\ln\beta\, +\,r_i(\beta)$, $i=1,2$, with $r_i(\beta)$ analytic for $\beta\ll 1$. This is done for YM models with gauge groups ${\mathrm SU}(N\not= 2)$. Of course,  it is desirable to prove that the model exists and the mass gaps persist in the continuum limit when the lattice spacing $a\searrow 0$. Also, it would be nice to obtain these results for larger values of $\beta$.

We conclude by stating that our results extend to spacetime dimension $d=3$ and that a multiplicity two mass gap is supported by simulations (see e.g. Ref. \cite{Ochs,Munster,Teper,Seo}).\vspace{5mm}

\noindent{\underline{ACKNOWLEDGEMENTS}}: We thank A. Cucchieri, J. Dimock, T. Pereira and E. Seiler for discussions.\vspace{3mm}
\appendix{\begin{center}{\bf APPENDIX A: Proof of the Elitzur's Theorem and Gauge Fixing}\end{center}}
\lb{appA}
\setcounter{equation}{0}
\setcounter{lemma}{0}
\setcounter{thm}{0}
\setcounter{rema}{0}
\renewcommand{\theequation}{A\arabic{equation}}
\renewcommand{\thethm}{A\arabic{thm}}
\renewcommand{\thelemma}{A\arabic{lemma}$\:$}
\renewcommand{\therema}{{\em{A\arabic{rema}$\:$}}}\vspace{.5cm}
In this Appendix, we first discuss gauge fixing. Then, we give the proof of our Elitzur´s theorem.

For concreteness and simplicity, we take the spacetime lattice to be in $d=2$ Euclidean dimensions and consider the gauge fixing procedure for one and two bond variables. The procedure generalizes to $d\geq 2$ and to a collection of bonds of a set of disjoint Cayley trees \cite{GJ,Gat}. Circuits with loops are {\em not} allowed. We first consider a system with only four neighboring plaquettes, which can be considered as a subset in the interior of the lattice $\Lambda$. Let $p_1,\,p_2,\,p_3,\,p_4$ denote these plaquettes. The numbers depicted in Figure 1 label the system positively (up and right directions are taken to be positive here) oriented bond variables. We will show that the value of the four-plaquette partition function $Z$ is unchanged if we set the bond variables $g_1$ and $g_5$ equal to the identity $1\in\mathcal G$. This is our gauge fixing! 

\tikzset{every picture/.style={line width=0.75pt}} 
\begin{center}
\begin{tikzpicture}[x=0.6pt,y=0.6pt,yscale=-1,xscale=1]\lb{fig1}
	
	\draw   (213,45.35) -- (313.78,45.35) -- (313.78,138.97) -- (213,138.97) -- cycle ;
	\draw   (313.78,45.35) -- (414.56,45.35) -- (414.56,138.97) -- (313.78,138.97) -- cycle ;
	\draw   (213,138.97) -- (313.78,138.97) -- (313.78,232.6) -- (213,232.6) -- cycle ;
	\draw   (313.78,138.97) -- (414.56,138.97) -- (414.56,232.6) -- (313.78,232.6) -- cycle ;
	
	\draw (261.27,122.11) node [anchor=north west][inner sep=0.75pt]  [font=\scriptsize] [align=left] {$\displaystyle 1$};
	\draw (303.02,87.33) node [anchor=north west][inner sep=0.75pt]  [font=\scriptsize] [align=left] {$\displaystyle 2$};
	\draw (257.39,51.45) node [anchor=north west][inner sep=0.75pt]  [font=\scriptsize]  {$3$};
	\draw (218.08,87.33) node [anchor=north west][inner sep=0.75pt]  [font=\scriptsize] [align=left] {$\displaystyle 4$};
	\draw (362.05,124.78) node [anchor=north west][inner sep=0.75pt]  [font=\scriptsize] [align=left] {$\displaystyle 5$};
	\draw (402.36,85.99) node [anchor=north west][inner sep=0.75pt]  [font=\scriptsize] [align=left] {$\displaystyle 6$};
	\draw (361.17,47.54) node [anchor=north west][inner sep=0.75pt]   [align=left] {{\scriptsize 7}};
	\draw (261.27,217.07) node [anchor=north west][inner sep=0.75pt]  [font=\scriptsize] [align=left] {$\displaystyle 8$};
	\draw (298.7,176.77) node [anchor=north west][inner sep=0.75pt]  [font=\scriptsize] [align=left] {9};
	\draw (220.18,176.77) node [anchor=north west][inner sep=0.75pt]  [font=\scriptsize] [align=left] {10};
	\draw (355.51,215.56) node [anchor=north west][inner sep=0.75pt]  [font=\scriptsize] [align=left] {11};
	\draw (395.82,178.11) node [anchor=north west][inner sep=0.75pt]  [font=\scriptsize] [align=left] {12};
	\draw (255,79) node [anchor=north west][inner sep=0.75pt]   [align=left] {$\displaystyle p_{1}$};
	\draw (353,81) node [anchor=north west][inner sep=0.75pt]   [align=left] {$\displaystyle p_{2}$};
	\draw (256,171) node [anchor=north west][inner sep=0.75pt]   [align=left] {$\displaystyle p_{3}$};
	\draw (353,169) node [anchor=north west][inner sep=0.75pt]   [align=left] {$\displaystyle p_{4}$};
\end{tikzpicture}\vspace{2mm}

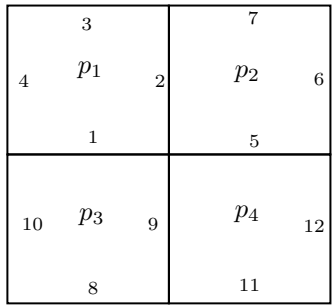
Figure 1. The four-plaquette system and its bond labels.\vspace{4mm}
\end{center}

Letting the four-plaquette system action be denoted by $A(\mathcal U)$, we also consider the expectation of functions $F_1$ and $F_5$ of the set $\mathcal U$ of all the twelve bond variables in the plaquettes $p_1,\,p_2,\,p_3,\,p_4$ 
\bequ\lb{Eli1}
\langle F_1(g_1)\,F_5(g_5)  \rangle\,=\,\dfrac1Z\;\dis\int_{\mathcal U}\;F_1(g_1)\,F_5(g_5)\,e^{-A(\mathcal U)}\,d\nu(\mathcal U)\,,
\eequ
For $d\nu({\mathcal U})\,=\,\prod_{g\in\mathcal U}\,d\mu(g)$, we show that
\bequ\lb{Eli2}
\langle F_1(g_1)\,F_5(g_5)  \rangle\,=\,\dis\int_\mathcal U\;F_1(g)\,d\mu(g) \;\dis\int_{\mathcal U}\;F_5(g)\,d\mu{(g)}\,\equiv\,\langle F_1\rangle_0\;\langle F_5\rangle_0\,,
\eequ
where, recalling the notation of section \ref{sec4}, the null label means the action $A(\mathcal U)$ has been set to zero in the expectation. This is a consequence of Elitzur´s theorem \cite{Gat,Eli}.

To talk about gauge fixing, it is important to recall that the normalized compact gauge group Haar measure obeys the left-right invariance property. For a single bond normalized Haar measure $d\sigma(U)$, this property is given as follows (see e.g. \cite{Bump,Simon2,Far}). Let $f(U)$ be a function of the bond variable $U\in\mathcal G$ and let $W\in{\mathcal G}$. Then,
\bequ\lb{lrinvariance}
\int_{\mathcal G}\,f(U)\,d\sigma(U)\,=\, \int_{\mathcal G}\,f(WU)\,d\sigma(U)\,=\, \int_{\mathcal G}\,f(UW)\,d\sigma(U)\,. 
\eequ
Here, as before, we take $\mathcal G\,=\,{\mathrm SU}(N)$ but $\mathcal G$ can be any compact Lie group.

Suppressing the $(1/g^2)$ pre-factor, we write schematically the four-plaquette system action as (here $A_i$ denotes the action of the plaquette $p_i$, $i=1,2,3,4$)
\bequ\lb{SA}
A(\mathcal U)\,=\,A_1\,+\,A_2\,+\,A_3\,+\,A_4\,=\,\tr\left[g_1g_2g_3^\dagger g_4^\dagger\,+\,g_5g_6g_7^\dagger g_2^\dagger\,+\,g_{8}g_{9}g_1^\dagger g_{10}^\dagger\,+\,g_{11}g_{12}g_5^\dagger g_{9}^\dagger\right]\,.
\eequ

The $g_1=1$ gauge fixing procedure consists of making a change of variables in $g_2$, $g_5$ and $g_{9}$ to eliminate $g_1$ in the action, after using Eq. (\ref{lrinvariance}) and the the cyclic property of the trace. With this, making the change of variables 
\bequ\lb{gauget}
\hat g_2\equiv g_1g_2\quad,\quad \hat g_5^\dagger\equiv g_5^\dagger g_1^\dagger\quad,\quad\hat g_{9}\equiv g_{9} g_1^\dagger\,,\eequ
the action of Eq. (\ref{SA}) becomes
\bequ\lb{SA2}
A(\mathcal U)\,=\,\tr\left[\hat g_2 g_3^\dagger g_4^\dagger\,+\,\hat g_5 g_6 g_7^\dagger \hat g_2^\dagger\,+\,g_8 \hat g_{9}^\dagger g_{10}^\dagger\,+\,g_{11}g_{12}\hat g_5^\dagger\hat g_{9}^\dagger\right].
\eequ
This is the same expression as the one in Eq. (\ref{SA}) upon setting $g_1=1$ and replacing $g_2$, $g_5$ and $g_{9}$ by their hat versions.
The hat variables are integration variables, the bond Haar measures are invariant by these changes of variables so that the partition function $Z$ remains unchanged.

Next, with $g_1=1$, we fix $g_5=1$. For this, consider two additional plaquettes, say the plaquettes $p_5$ and $p_6$ laying at the right of $p_2$ and $p_4$, respectively. We have now a $2\times 3$, six-plaquette system. We can follow the same procedure as before to fix $g_5=1$ in the action $A(\mathcal U)\,=\,A_1\,+\,A_2\,+\,A_3\,+\,A_4\,+\,A_5\,+\,A_6$, for the new set $\mathcal U$.

We now turn to Elitzur´s theorem and prove Eq. (\ref{Eli2}) in the context of the four plaquette system given in Figure 1. Besides the previously defined notation [see Eq. (\ref{Eli1})], here we also use the following notation. If the directed bond labeled by $1$ in the figure starts at the source point $x\in Z_t^\infty$ and ends at $y$, we denote $x\equiv 1_-$ and $y\equiv 1_+$ and we let $g_1\in\mathcal G$ to be the corresponding bond variable. Let $A(g)$ denote the Wilson action of Eq. (\ref{SA}) where we only display explicitly the dependence on the bond $1$. Also, consider the set of functions
$$
h\,=\,\left\{ h_x\equiv h(x)\in\mathcal G, x\in Z_t^\infty\right\}\,,
$$
describing a global gauge group. Then, the change of variables $\hat g_1\,=\,h_{1-}\,g_1\,h^\dagger_{1+}$ is a gauge transformation and we have the equality between the actions $A(\hat g_1)\,=\,A(g_1)$ and the normalized Haar measures $d\mu(\hat g_1)\,=\,d\mu(g_1)$. Hence, with the normalized Haar measure $d\rho(h)\,=\,\prod_{x\in\mathcal U}\,d\mu(h_x)$, we can write the average of a function $F_1$ of $g_1$ as
\bequ\lb{avg1}\barr{lll}
\langle F_1(g_1)\rangle&\equiv&\dfrac1Z\,\dis\int\,F_1(g_1)\,e^{-A(g_1)}\;d\nu(\mathcal U)\vspace{1mm}\\
&=& \dis\int d\rho(h)\;\dfrac1Z\,\dis\int\,F_1(g_1)\,e^{-A(g)}\;d\nu(\mathcal U)\vspace{1mm}\\
&=&\dfrac1Z\,\dis\int d\mu(h_{1-})\,d\mu(h_{1+}) \,F_1(h_{1-}^\dagger\,\hat g_1\,h_{1+})\,e^{-A(\hat g)}\;d\nu(\hat {\mathcal U})\,.
\earr\eequ 
Here, $d\nu(\hat{\mathcal U})$ denotes the four-plaquette product of twelve bond normalized Haar measures where we made the change of variable $g_b\,=\,h_{1-}^\dagger\,\hat g_b\,h_{1+}$, for any $b\in\mathcal U$. We also used the equality $\int d\rho(h)\,=\,1$. Moreover, using Eq. (\ref{lrinvariance}) in Eq. (\ref{avg1}), we obtain
\bequ\lb{avg2}
\langle F(g_1)\rangle\,=\,\dis\int\,F(h_{1+})\,d\mu(h_{1+})\,.
\eequ

We now consider also bond $5$ in Figure 1. Repeating the same arguments, we have
\bequ\lb{avg3}
\langle F(g_1)\;F_5(g_5)\rangle\,=\,\dfrac1Z\,\dis\int\,F_1(h^\dagger _{1-}\,\hat g_1\,h_{1+})\;F_5(h^\dagger _{5-}\,\hat g_5\,h_{5+})\,e^{-A(\hat g_1,\hat g_5)}\;d\nu(\hat{\mathcal U})\,d\rho(h)\,.
\eequ
Performing the $h_{5+}$ and then the $h_{1+}$ integrals and using Eq. (\ref{lrinvariance}), we get  Eq. (\ref{Eli2}).

It is important to emphasize that, of course, the above procedure works as well when the four plaquettes are a improper subset e.g. the entire lattice $\Lambda$ so that the set $\mathcal U$ is the set of all bonds in $\Lambda$.

In closing this Appendix, we explain how to generalize the gauge fixing procedure to spacetime dimension $d\geq 2$. Here, $(2d-1)$ plaquettes share e.g. the bond $g_1$ and we consider these plaquettes. Let $k_-$ and $k_+$ be the initial and terminal points of the bond $b_k$, respectively. To eliminate the bond variable $g_1$ of the bond $b_1$, we make the changes of variables in the $[2(d-1)+1]$ bond variables, other than $g_1$, that have the point $1_+$ as their initial or terminal point. This fixes $g_1=1$.

To fix $g_5=1$, we proceed as before considering the $[2(d-1)]$ plaquettes that have $g_5$ in common. For bonds on the boundary of the lattice, the preceding gauge fixing procedure can be adapted.

We end by remarking that the change of variables in the procedure to fix $g_1=1$ [see Eq. (\ref{gauget})] is a gauge transformation $\hat g_b\,=\,h_{b-}\,g_b\,h^\dagger_{b+}$, where the global gauge $h\,=\,\left\{h_x\,,\,x\in Z_t^\infty \right\}$
element is $h_{2-}\,=\,h_{10+}\,=\,h_{5-}\,=\,g_1$ (here we have $x=2-\,=\,5-\,=\,10+)$. For all other $x\in\Lambda$, we have $h_x\,=\,1$.\vspace{3mm}
\appendix{\begin{center}{\bf APPENDIX B: Wilson Action in the Gluon Field Parametrization}\end{center}}
\lb{appB}
\setcounter{equation}{0}
\setcounter{lemma}{0}
\setcounter{thm}{0}
\setcounter{rema}{0}
\renewcommand{\theequation}{B\arabic{equation}}
\renewcommand{\thethm}{B\arabic{thm}}
\renewcommand{\thelemma}{B\arabic{lemma}$\:$}
\renewcommand{\therema}{{\em{B\arabic{rema}$\:$}}}\vspace{.5cm}
Here, we give an expansion for the Wilson action in terms of the gluon fields in the case of the gauge group $\mathrm{SU}(2)$. This expansion, together with a deeper understanding of the contents of the Haar measure in terms of the gauge fields, are important points to obtain the solution to mass gap problem in the continuum limit $a\searrow 0$. It also sets the stage for a renormalization group approach to determine the low-lying energy-momentum spectrum. With some adjustments, our treatment can be generalized to the gauge group ${\mathrm SU}(N\not= 2)$.

Our expansion is given up to and including the fourth order in the fields. The terms are classified according to the Renormalization Group formalism point of view (see Refs. \cite{Bal,Bal2,Dim,Dim2,Dim3,Dim4}), with canonical scaling in spacetime dimension $d=4$.

We write an element of the group $\mathrm{SU}(2)$ as $\exp[i\vec B.\vec \sigma]$, where $\vec \sigma\,=\, (\sigma_1,\sigma_2,\sigma_3)$ and $\sigma_{i=1,2,3}$ are the hermitian and traceless Pauli matrices. The well known identity
$$
(\vec A\cdot\vec\sigma)\,(\vec B\cdot\vec\sigma)\,=\,\vec A\cdot\vec B\,+\,i\,\vec\sigma\cdot (\vec A\times \vec B)\,.
$$
is repeatedly used, with a three-dimensional vector notation for the elements of the Lie algebra of $\mathcal G\,=\,{\mathrm SU}(2)$.

For the plaquette $p=p_{\mu\nu}$ in the $\mu\nu$-plane (with $\mu<\nu$), we write
$$
A_p\,=\,2\,\Re\tr (1\,-\,U_p)\,=\,2\,\tr (1\,-\,U_p)\,,
$$
where 
$$
U_p\,=\,\;\left[ \exp\left(i\vec A_\mu(x)\cdot\vec\sigma\right)\,\exp\left(i\vec A_\nu(x^+_\mu)\cdot\vec\sigma\right) \right] \;
\left[ \exp\left(i\vec A_\nu(x)\cdot\vec \sigma\right)\,\exp\left(i\vec A_\mu(x^+_\nu)\cdot\vec\sigma\right)\right]^\dagger\,.
$$

Using the simplified notation $A_\mu\equiv\vec A_\mu(x)$, $A_\nu^+\equiv\vec A_\nu(x^+_\mu)$, $A_\nu\equiv\vec A_\nu(x)$ and 
$A_\mu^+\equiv\vec A_\mu(x^+_\mu)$, we 
have
\bequ\lb{UUU}
U_p\,=\, \left[e^{X_1}\,e^{Y_1}\,\equiv\,e^{Z_{12}}\right]\;
\left[e^{X_2}\,e^{Y_2}\,\equiv\,e^{Z_{43}}\right]^\dagger\,=\,e^{Z_{12}}\;e^{Z_{43}^\dagger}\,=\,e^{Z_p}\,.
\eequ
where $Z_p\,=\,i \vec K\cdot\vec\sigma$.

Now, writing 
$$
1\,-\,U_p\,=\, -i\vec\sigma\cdot\vec K\,+\,\dfrac12\,\vec K\cdot\vec K \,+\,\dfrac i{3!} \left(  \sigma\cdot\vec K\right)\,|\vec K|^2\,-\,\dfrac1{4!} \, |\vec K|^4\,,
$$
we get
\bequ\lb{AK}
\dfrac12\,A_p\,=\, \tr (1\,-\,U_p)\,=\, |\vec K|^2\,-\, \dfrac1{12}\,|\vec K|^4\,+\,\mathcal O(5)\,,
\eequ
where $\mathcal O(5)$ means terms involving at least five $\vec K$ factors.

Now, we use the Baker-Campbell-Hausdorff formula \cite{Far,Sternberg,Moore} to write $e^X\,e^Y\,=\,e^Z$ [check Eq. (\ref{BCH})] to obtain
$$
Z\,=\,X\,+\,Y\,+\,\dfrac12\,[X,Y]\,+\,\dfrac1{12}\,[X,[X,Y]]\,+\,\dfrac1{12}\,[Y,[Y,X]]\,+\,\dfrac1{24}\left[ Y^2\,,\,X^2\right]\,+\,\mathcal O(5)\,,
$$
where $\mathcal O(5)$ means terms with five or more $X$ and/or $Y$ factors, in each factor of $U_p$ in Eq. (\ref{UUU}). Doing this, we obtain
$$\barr{lll}
Z_{12}&=& i\vec \sigma\,\cdot\, \left[ \vec A_\mu\,+\,\vec A_\nu^+\,-\,\vec (A_\mu\times\vec A_\nu^+) \right]\,+\,  \dfrac13\,\vec A_\mu\times (\vec A_\mu\times\vec A_\nu^+)\,+\,\dfrac13\,\vec A_\nu^+\times (\vec A_\nu^+\times\vec A_\mu)\,+\,\mathcal O(5)     \vspace{2mm}\\
&\equiv&i\vec\sigma\,\cdot\,\vec C_{12}\,+\,\mathcal O(5)\,,
\earr
$$
and
$$ \barr{lll}
Z_{43}^\dagger&=&-\,i\vec \sigma\,\cdot\, \left[ \vec A_\mu\,+\,\vec A_\nu\,-\,\vec A_\nu\times\vec A_\mu^+ \right]\,+\,  \dfrac13\,\vec A_\nu\times (\vec A_\nu\times\vec A_\mu^+)\,+\,\dfrac13\,\vec A_\mu^+\times (\vec A_\mu^+\times\vec A_\nu)\,+\,\mathcal O(5)     \vspace{2mm}\\
&\equiv&i\vec\sigma\,\cdot\,\vec C_{43}\,+\,\mathcal O(5)\,.
\earr $$
Hence, applying again the Baker-Campbell-Hausdorff formula to the last equality of Eq. (\ref{UUU}), it follows that
$$\begin{array}{lll}
Z_p&=& \vec\sigma.\left[\vec C_{12}\,+\,\vec C_{43}\,-\,\vec C_{12}\cdot\vec C_{43}\,+\,\dfrac13\,\vec C_{12}\times\left(\vec C_{12}\,+\,\vec C_{43}\right) \,+\,\dfrac13\,\vec C_{43}\times\left(\vec C_{12}\,+\,\vec C_{43}\right)\right]\,+\,\mathcal O(4) \vspace{2mm}\\
&=& i\,\vec\sigma\cdot\vec K\,+\,\mathcal O(4)\,.                    
\end{array}     
$$

Now, writing
$$
\vec K\,=\, \vec C_{12}\,+\,\vec C_{43}\,-\,\vec C_{12}\times\vec C_{43}\,+\,\vec R_c\,+\,\mathcal O(4)\,,
$$
where $\vec R_c$ are the cubic terms only, we have
$$
\vec C_{12}\,=\,\left(\vec A_\mu\,+\,\vec A_\nu^+\right)\,+\,\left(\vec A_\nu^+\times \vec A_\mu \right)\,\equiv\, \vec L_{12}\,+\,\vec R_{12} 
$$
and 
$$
\vec C_{43}\,=\,-\,\left(\vec A_\nu\,+\,\vec A_\mu^+\right)\,+\,\left(\vec A_\mu^+\times \vec A_\nu \right)\,\equiv\, \vec L_{43}\,+\,\vec R_{43}\,, 
$$
and with
$$\barr{lll}
\vec R_c&=&\dfrac13\,\vec L_{12}\times\left(\vec L_{12}\times\vec L_{43}\right)\,+\,\dfrac13\,\vec L_{43}\times\left(\vec L_{43}\times\vec L_{12}\right)\vspace{2mm}\\
&=&\dfrac13\, \left(\vec A_\mu\,+\,\vec A_\mu^+\,+\,\vec A_\nu\,+\,\vec A_\nu^+  \right)\times\left[\left(\vec A_\mu\,+\,\vec A_\nu\right)\times\left(\vec A_\mu^+\,+\,\vec A_\nu^+\right)   \right]\,.\earr
$$

Next, we set
$$
\vec C_{12}\,+\,\vec C_{43}\,=\,\vec L_{\mu\nu}\,+\,\vec R_{\mu\nu}\,+\,\vec T_{\mu\nu}\,,
$$
where
\bequ\lb{tensors}\barr{lll}
\vec L_{\mu\nu}\,=\,(\vec A_\mu\,-\,\vec A_\mu^+)\,-\,(\vec A_\nu\,-\,\vec A_\nu^+)\,=\,\partial^a_\mu \vec A_\nu\,-\,\partial^a_\nu \vec A_\mu\,,
\vspace{2mm}\\
\vec R_{\mu\nu}\,=\,(\vec A_\nu^+\,\times\,\vec A_\mu)\,+\,(\vec A_\nu\,\times\,\vec A_\mu^+)
\vspace{2mm}\\
\vec T_{\mu\nu}\,=\,\dfrac13\,\left[ \vec A_\mu\,\times\,(\vec A_\mu\,\times\,\vec A_\nu^+) \,+\,\vec A_\nu^+\,\times\,(\vec A_\nu^+\,\times\,\vec A_\mu^+)\,+\,\vec A_\nu\,\times\,(\vec A_\mu^+\,\times\,\vec A_\nu)\,+\,\vec A_\mu^+\,\times\,(\vec A_\nu\,\times\,\vec A_\mu^+)  \right]\,.
\earr\eequ
Thus, using the above equations, after some algebra, we obtain
\bequ\lb{K2}
|\vec K|^2\,=\,\vec K\cdot\vec K\,=\,|\vec L_{\mu\nu}|^2\,+\,|\vec R_{\mu\nu}|^2\,+\,2\,\vec L_{\mu\nu}\cdot\left(\vec R_{\mu\nu}\,+\,\vec T_{\mu\nu}\right)\,+\, | (\vec A_\mu\,+\,\vec A_\nu^+)\,\times\,(\vec A_\mu^+\,+\,\vec A_\nu)   |^2\,+\,\mathcal O(5)\,.
\eequ

Recalling that Eq. (\ref{K2}) is to be replaced in Eq. (\ref{AK}), we obtain the action $A_p$ written in terms of the tensors $\vec L_{\mu\nu}$, $\vec R_{\mu\nu}$ and $\vec T_{\mu\nu}$. Namely, we have
\bequ\lb{endK}
\dfrac12\,A_p\,=\,|\vec L_{\mu\nu}|^2\,\,+\,|\vec R_{\mu\nu}|^2\,+\,2\,\vec L_{\mu\nu}\cdot\left(\vec R_{\mu\nu}\,+\,\vec T_{\mu\nu}\right)\,+\,| (\vec A_\mu\,+\,\vec A_\nu^+)\,\times\,(\vec A_\mu^+\,+\,\vec A_\nu)|^2\,+\,\mathcal O(5)\,.
\eequ
In the local potential approximation  $A_p$ agrees with the naive approximation with $F^a_{\mu\nu}(x)$ given in Eq. (\ref{fmunua}).

Note that, from the renormalization group point of view, with canonical scaling, and considering spacetime dimension $d=4$, the terms in Eqs. (\ref{K2}) and (\ref{endK}) are all marginal, except for the $\left[\vec L_{\mu\nu}\cdot\left(\vec R_{\mu\nu}\,+\,\vec T_{\mu\nu}\right)\right]$ and $|\vec L_{\mu\nu}|^4$ which are irrelevant. There is nothing like a usual relevant quadratic mass term.

The absence of a relevant mass term in the Wilson action must be taken into account in proving the existence of a mass gap, in the continuum limit $a\searrow 0$ of YM. A mass term may arise from elsewhere. A possibility is that it comes from the Haar measure density. The Haar measure for ${\mathrm SU}(2)$ is given in Eq. (\ref{haarsu2}) and it does exhibit a local positive mass term in the exponentiated density. This is why in the next appendix we analyze the Haar measure using the gluon field parametrization. After beginning with a general treatment, we analyze the gauge group ${\mathrm SU}(2)$ in more detail. \vspace{3mm}
\appendix{\begin{center}{\bf APPENDIX C: ${\mathrm SU}(N)$ Haar Measure in the Gluon Field Parametrization}\end{center}}
\lb{appC}
\setcounter{equation}{0}
\setcounter{lemma}{0}
\setcounter{thm}{0}
\setcounter{rema}{0}
\renewcommand{\theequation}{C\arabic{equation}}
\renewcommand{\thethm}{C\arabic{thm}}
\renewcommand{\thelemma}{C\arabic{lemma}$\:$}
\renewcommand{\therema}{{\em{C\arabic{rema}$\:$}}}\vspace{.5cm}
In this appendix, we determine the Haar measure in the gluon field parametrization. Although the Haar measure for the ${\mathrm SU}(N)$ group has been written in other parametrizations (Euler angles) \cite{Molin,Nelson,Holland,Holland2,Marinov,Marinov2,Euler,Euler2,Euler3} (see also the review paper \cite{EULER} and references therein), we never found the ${\mathrm SU}(N)$ Haar measure given in terms of the, more physical, gluon fields.

For ${\mathrm SU}(2)$, in Eq. (\ref{haarsu2}), we give a closed form expression of the Haar measure in terms of Riemann zeta function. It involves a series with positive coefficients which are related to Bernoulli numbers.  The exponential of the measure density gives a mass term.

A deep question is that if this property is shared by the group ${\mathrm SU}(N\not= 2)$. Below, using the Cartan-Maurer form \cite{Sternberg,Sternberg2}, we obtain an entire expansion in terms of the gluon fields but we have not determined the sign of the non-quadratic terms. The question arises on how this property of existence of a mass term is related to the geometry of gauge group. 

We need the density with respect to the Lebesgue measure for the left invariant Haar measure (hence, right invariant Haar measure since ${\mathrm SU}(N)$ is compact). For group elements $g$ and $h$ near the identity, we write $g\,=\,\exp(i\vec a.\vec \theta)$ and $h\,=\,\exp(i\vec b.\vec \theta)$. Then, the left multiplication by $h$ is given by the product
$$
h\,g\,=\,\exp(i\vec b.\vec \theta)\, \exp(i\vec a.\vec \theta)\,\equiv\,\exp(i\vec c.\vec \theta)\,,
$$
with $\vec c\,=\,\vec c(\vec a, \vec b)$. Thus, the density $\rho_h(\vec b)$ is the pullback of the volume form (see e.g. Refs. \cite{Simon2,Far})
\bequ\lb{pull}
\rho_h(\vec b)\,=\, \dfrac{{\mathrm const}}{\det J}\,=\, {\mathrm const}\: \;e^{-(-\ln\rho_h(\vec b))}\,,
\eequ
where  ${\mathrm const}$ is a normalization constant and $J$ is the Jacobian matrix 
$$ J\,=\,\left(
\begin{array}{lcl} 
	\dfrac{\partial c_1}{\partial b_1}&\:\ldots\:&\dfrac{\partial c_1}{\partial b_n}\vspace{2mm}
	\\\vdots&\ddots&\vdots\vspace{2mm}\\
	\dfrac{\partial c_n}{\partial b_1}&\:\ldots\:&\dfrac{\partial c_n}{\partial b_n}     
\end{array} \right)_{|_{\vec c\,=\,0}}\,,
$$
where, here, $n$ denotes the Lie algebra dimension $n\,\equiv\,\delta_N\,=\, N^2-1$.
The formula given in Eq. (\ref{pull}) is valid at least in a neighborhood of the identity.

For $\vec c(\vec a,\vec b)$, we can use again the Baker-Campbell-Hausdorff formula (see Refs. \cite{GJ,Far,Moore,Sternberg})
$$
e^C\,=\,e^A\,e^B\,,
$$
where, for small $A$ and $B$, we have
\bequ\lb{BCH}
C\,=\,A\,+\,B\,+\,\dfrac12\,[A,B]\,+\,\dfrac1{12}[A,[A,B]]\,+\,\dfrac1{12}[B,[B,A]]\,+\,\dfrac1{24}\,[B^2,A^2]\,\,,
\eequ
and the remainder is of total order five in the $A$´s and $B$´s.

Note that in $J$ we only need the linear terms in $B$ (see \cite{Moore} for the simplification in this case). We find that, to order $|\vec b|^2$
$$
\det J\,=\, 1\,+\,\dfrac 16\, (\vec b, C_2 \vec b)\,+\,{\mathcal O}(|\vec b|^3)\,,
$$
with 
$$ (C_2)_{ij}\,=\,f_{ik\ell}f_{jk\ell}\,=\,N\,\delta{ij}\,. $$
Here $f_{ik\ell}$ are the completely antisymmetric structure constants of ${\mathrm SU}(N)$ and $C_2$ is the quadratic Casimir matrix in the adjoint representation of the Lie algebra of the group, i.e. $C_2\,=\, N\,I$, with $I$ being the identity (see \cite{Haber}).

The above representation for the Haar measure density has the defect that the smoothness in $\vec b$ is only evident for small $|\vec b|$. Alternatively, we can use another well-known construction which shows that the density is real analytic in $\vec b$.  We give a brief summary of this construction below.

Roughly speaking the construction uses the Maurer-Cartan invariant one-form \cite{Sternberg,Sternberg2}. Wedge products are taken to obtain the invariant Haar measure. Explicitly, writing $U\,=\, \exp[i\vec b.\vec \theta]\,\in\, {\mathrm SU}(N)$. Then, $w\,=\,U^{-1}\,dU$ is a matrix one-form. The real and imaginary element are left-invariant one-forms. To see this, consider the left multiplication of $U$ by an arbitrary fixed element of the group
$$
V^{-1}\,dU\,=\, U^{-1} \exp(-i\vec a.\vec\theta)\,d(\exp(i\vec a\vec\theta)\, U\,=\,U^{-1}\,dU\,.
$$
Using the representation for the differential
\bequ\lb{integs}
d\left( e^{B(t)} \right)\,=\,\dis\int_0^1\,e^{(1-s)B(t)}\,dB(t)\,e^{sB(t)}\,ds\,,
\eequ
we have 
$$
w\,\equiv\, U^{-1}\,dU\,=\,i\,\dis\int_0^1\,e^{-i\,t\,\vec b.\vec\theta}\,d\vec b.\vec\theta\,e^{i\,t\,\vec b.\vec\theta}\,.
$$

From this formula, we see that $w$ is a matrix of one-forms. The real and imaginary parts of each matrix elements are left-invariant one-forms which are real analytic in $\vec b$.

Taking the wedge product (with $(N^2-1)$ factors)
$$
w_{N^2-1}\,=\,w^\#\,\wedge\,w^\#\,\wedge\,\ldots\,\wedge w^\#
$$
where $w^\#$ is the real or imaginary part of $w$, gives us, up to a normalization constant, a matrix for Haar measure $(N^2-1)$-forms of $w_{N^2-1}$ which is not zero. The coefficient of $db_1\wedge\ldots\wedge db_{N^2-1}$ gives the density.

For example, for ${\mathrm SU}(2)$ at $\vec b=0$, $w\,=\,id\vec b\,\cdot\vec\theta$, if we take
$$
w_3\,=\,\Im w\,\wedge\,\Im w\,\wedge\,\Re w\,,
$$
then $w_3$, at $\vec b=0$, is
given by
\bequ
\barr{lll}\lb{wedges}
w_3&=&\left(\barr{cc}db_3&db_1\vspace{1mm}\\db_1&-db_3  \earr\right)\,\wedge\,\left(\barr{cc} db_3&db_1\vspace{1mm}\\db_1&-db_3 \earr\right)\,\wedge\,\left(\barr{cc}0&db_2\vspace{1mm}\\  -db_2&0\earr\right)\vspace{4mm}\\
&=&-2\;\left(\barr{cc}db_1\wedge db_2 \wedge db_3&0\vspace{2mm}\\
0&  db_1\wedge db_2 \wedge db_3\earr\right)\,\not=0\earr\,.
\eequ

Computing in detail the integrals like the one given in Eq. (\ref{integs}), we shall obtain the Haar measure given in Eq. (\ref{haarsu2}). For ${\mathrm SU}(N)$, the equation similar to Eq. (\ref{wedges}) involves $(N^2-1)$ wedge products!


\vspace{2cm}
\begin{thebibliography}{8fdw}
\bibitem{YMOld} C.N. Yang and R.L. Mills: {\em Conservation of Isotopic Spin and Isotopic Gauge Invariance}, Phys. Rev. 96, 191-195 (1954).
\bibitem{FGL}H. Fritsch, M. Gell-Mann and H. Leutwyler, {\em Advantages of the Color Octet Gluon Picture}, Phys. Lett. {B47}, 365 (1973).
\bibitem{Wil2} K. Wilson, {\em Confinement of Quarks}, Phys. Rev. {D10}, 2445 (1974).
\bibitem{Wil} K. Wilson, in {\em New Phenomena in Subnuclear
	Physics}, Part A, A. Zichichi ed. (Plenum Press, NY, 1977).
\bibitem{Banks} T. Banks et al., {\em Strong-coupling Calculations of the Hadron Spectrum of Quantum Chromodynamics}, Phys. Rev. {D15}, 1111 (1977).
\bibitem{FK} J. Fr\"ohlich and C. King, {\em Meson Masses and the U(1) Problem in Lattice QCD}, Nucl. Phys { B290}, 157 (1987).
\bibitem{Mon} I. Montvay, {\em Numerical Calculation of Hadron Masses in Quantum Chromodynamics}, Rev. Mod. Phys. {59}, 263 (1987).
\bibitem{MW} F. Myhrer and J. Wroldsen, {\em The Nucleon-Nucleon Force and the Quark Degrees of Freedom}, Rev. Mod. Phys. {60},
629 (1988).
\bibitem{Schreiber} D. Schreiber, {\em t Expansion of QCD Baryons}, Phys. Rev. D48, 5393 (1993).
\bibitem{Mach} R. Machleidt, {\em The Nuclear Force in the Third Millennium}, Nucl. Phys. A689, 11 (2001).
\bibitem{Mach2} R. Machleidt, K. Holinde, and Ch. Elster, {\em The Bonn Meson-exchange Model for the Nucleon—nucleon Interaction} Phys. Rep.
{149}, 1 (1987).
\bibitem{Creu} M. Creutz, {\em Lattice gauge theory: A retrospective}, Nucl. Phys. (Proc. Suppl.) 94, 219 (2001).
\bibitem{Creu2} M. Creutz, {\em Quarks, Gluons and Lattices}
(Cambridge University Press, Cambridge,  1983). 
\bibitem{MM} I. Montvay and G. M\"unster, {\em Quantum Fields on a
	Lattice} (Cambridge University Press, Cambridge, 1997).
\bibitem{GJ} J. Glimm and A. Jaffe, {\em Quantum Physics: A Functional
	Integral Point of View} (Springer Verlag, New York, 1986).
\bibitem{Sp} T. Spencer: {\em The Decay of the Bethe-Salpeter Kernel in $P(\phi)_2$ Quantum Field Models}, Commun. Math. Phys. 44, 143-175 (1975).
\bibitem{Si} B. Simon, {\em Statistical Mechanics of Lattice Models}
(Princeton University Press, Princeton, 1994).	
\bibitem{Sei} E. Seiler, Lect. Notes in Phys. {159}, {\em Gauge Theories as a Problem of Constructive Quantum Field Theory and Statistical Mechanics}
(Springer, New York, 1982).
\bibitem{OS} K. Osterwalder and E. Seiler, {\em Gauge Field Theories on a Lattice}, Ann. Phys. (NY) 110, 440 (1978).
\bibitem{QCD} P.A. Faria da Veiga, M. O'Carroll, and R. Schor, {\em Understanding baryons from first principles},  Phys.
Rev. D67, 017501 (2003).
\bibitem{2baryon}P.A. Faria da Veiga, M. O'Carroll, and R. Schor,
Phys. Rev. {D68}, 037501 (2003).
\bibitem{CMP} P.A. Faria da Veiga, M. O´Carroll and R. Schor: {\em Existence of Baryons, Baryon Spectrum and Mass Splitting in Strong Coupling Lattice QCD}, Commun. Math. Phys. 245, 383–405 (2004).
\bibitem{JMP} A. Francisco Neto, P.A. Faria da Veiga, and M. O'Carroll, {\em Existence of mesons and mass splitting in strong coupling lattice quantum chromodynamics }, J. Math. Phys. 45, 628 (2004).
\bibitem{2meson} P.A. Faria da Veiga, M. O'Carroll, and A.
Francisco Neto, {\em Meson-meson Bound States in a (2 +1)-Dimensional Strongly Coupled Lattice QCD Model}, Phys. Rev. D69, 097501 (2004).
\bibitem{2flavor2baryon}P.A. Faria da Veiga and M. O'Carroll, {\em Baryon-baryon bound state in a 2 +1 lattice QCD model with two flavors and strong coupling}, Phys. Rev. {D71}, 017503 (2005).
\bibitem{2flavor2meson} P.A. Faria da Veiga, M. O'Carroll, and A.
Francisco Neto, {\em Meson-meson Bound State in a 2+1 Lattice QCD Model with Two Flavors and Strong Coupling}, Phys. Rev. D72, 034507 (2005).
\bibitem{physlettb} P.A. Faria da Veiga and M. O'Carroll, {\em Baryon–baryon Bound states from First Principles in Lattice QCD with Two flavors and Strong Coupling}, Phys. Lett. {B643}, 109 (2006).
\bibitem{longo}P.A. Faria da Veiga and M. O'Carroll, {\em Baryon-baryon Bound States in Strongly Coupled Lattice QCD}, Phys. Rev. {D75}, 074503 (2007).
\bibitem{8fdw} P.A. Faria da Veiga and M. O'Carroll, {\em Eightfold Way from Dynamical First Principles in Strongly Coupled Lattice Quantum Chromodynamics}, J. Math. Phys. 49, 042303 (2008).
\bibitem{8fdwm} A. Francisco Neto, M. O’Carroll and P.A. Faria da Veiga: {\em Mesonic Eightfold Way from Dynamics and Confinement in Strongly Coupled Lattice Quantum Chromodynamics}, J. Math. Phys. 49, 072301 (2008).
\bibitem{PMJ1} P.A. Faria da Veiga, M. O'Carroll and J.C.V. Alvites: {\em One-baryon Spectrum and Analytical Properties of One-baryon Dispersion Curves in 3+1 Dimensional Strongly Coupled Lattice QCD with Three Flavors}, J. Math. Phys. 57, 032303 (2016).
\bibitem{PMJ2} P.A. Faria da Veiga, M. O'Carroll and J.C.V. Alvites: {\em On the Energy-momentum Spectrum and One-meson Dispersion Curves in (3+1)-Dimensional Strongly Coupled Lattice QCD with Three Flavors}, Rep. Math.Phys. 83, 207-242  (2019).
\bibitem{num1} H.R. Fiebig, H. Markum, A. Mih\'aly, and K. Rabitsch, {\em Forces between composite particles in QCD}, Nucl. Phys. Proc. Suppl. 53, 804 (1997).
\bibitem{num2} C. Stewart and R. Koniuk, {\em Hadronic Molecules in Lattice QCD}, Phys. Rev. D57, 5581 (1998).
\bibitem{num3} H.R. Fiebig and H. Markum, in {\em International Review of Nuclear Physics},  {\em Hadronic Physics from Lattice QCD}, A.M. Green ed., World Scientific, Singapoure (2003).
\bibitem{num4} Ph. de Forcrand and S. Kim, {\em The Spectrum of Lattice QCD with Staggered Fermions at Strong Coupling}, Phys. Lett. B645, 339 (2007).
\bibitem{Ochs} W. Ochs, {\em The Status of Glueballs}, J. Phys. G: Nucl. Part. Phys. 40, 043001 (2013).
\bibitem{Munster} G. M\"unster, {\em Strong Coupling Expansions for the Mass Gap in Lattice Gauge Theories}, Nucl. Phys. B190, 439-453 (1981).
\bibitem{Teper} A. Athenodorou and M. Teper, {\em ${\mathrm SU}(N)$ Gauge Theories in 2+1 Dimensions: Glueball Spectra and k-String Tensions},  J. High Energ. Phys. 2017, 15 (2017); idem  {\em The Glueball Spectrum of SU(3) Gauge Theory in $3+1$ Dimensions}, J. High Energ. Phys. 2020, 172 (2020).; P. Conkey, S. Dubovskya and M. Teper, {\em Glueball Spins in $D = 3$ Yang-Mills}, J. High Energ. Phys. 2019, 175 (2019).
\bibitem{Seo} K. Seo, {\em Glueball Mass Estimate by Strong Coupling Expansion in Lattice Gauge Theories}, Nucl. Phys. B209, 200- 216 (1982); K. Seo and A.  Ukawa, {\em Off-Axis Plaquette-Plaquette Correlation Functions}, Phys. Lett. B114, 329-332 (1982).
\bibitem{Schor1} R.S. Schor, {\em Existence of Glueballs in Strongly Coupled Lattice Gauge Theories} Nucl. Phys. {B222}, 71-82 (1983).
\bibitem{Schor2} R.S. Schor, {\em The Energy Momentum Spectrum of Strongly Coupled Lattice Gauge Theories}, Nucl. Phys.{B231}, 321-334 (1984).
\bibitem{Schor3} R.S. Schor, {\em Glueball Spectroscopy in Strongly Couples Lattice Gauge Theories}, Commun.Math.Phys. 92,  369-395 (1984).
\bibitem{Stein} E.M. Stein, {\em Harmonic Analysis}, Princeton University Press, Princeton, NJ (1993).
\bibitem{ROMP} P.A. Faria da Veiga and M. O'Carroll, {\em On Yang-Mills Stability Bounds and Plaquette Field Generating Function}, Rep. Math. Phys. 95, 303-380 (2025).
\bibitem{Gat} C. Gattringer, C.B. Lang, {\em Quantum Chromodynamics on the Lattice, An Introductory Presentation}, Lecture Notes in Physics 788, Springer, New York, 2010.
\bibitem{Far} J. Faraut, {\em Analysis on Lie Groups: An Introduction}, Cambridge University Press, Cambridge UK, 2008.
\bibitem{Moore} G.W. Moore, {\em Lecture Notes of Spring Course}, Rutgers University (2023). Check the source text in the  website https://www.physics.rutgers.edu/~gmoore/618Spring2018/GTLect8-LieGroupTheory-2018.pdf
\bibitem{Sternberg} S. Sternberg, {\em Lectures on Lie Algebras}, https://people.math.harvard.edu/~shlomo/ , (2004).
\bibitem{Hairer} A. Chandra, I. Chevyrev, M. Hairer and H. Shen, {\em Stochastic Quantisation of Yang-Mills-Higgs in 3D}, {\it Invent. Math.} (2024). https://doi.org/10.1007/s00222-024-01264-2.
\bibitem{Bump} D. Bump, {\em Lie Groups}, GTM225 Series, Springer, New York, 2000.
\bibitem{OS1} K. Osterwalder and R. Schrader, {\em Axioms for Euclidean Green's functions I}, Commun. Math. Phys. 31, 83-112 (1973).
\bibitem{OS2} K. Osterwalder and R. Schrader, {\em Axioms for Euclidean Green's functions. II}, Commun. Math. Phys. 42, 281-305 (1975).
\bibitem{RS1} M. Reed and B. Simon, {\em Modern Methods of Mathematical Physics}, vol. 1, {\sl Functional Analysis } (Academic Press, New York, 1972).
\bibitem{JR} J.M. Jauch and  F. Rohrlich, {\em The Theory of Photons and Electrons}, Springer-Verlag, New York (1976).
\bibitem{PMR} R. Mistry, P.A. Faria da Veiga and M. O´Carroll, {\em The Low-lying Energy-Momentum Spectrum for Unit Lattice YM Models in the Strong Coupling Regime}, manuscript in preparation.
\bibitem{Eli} S. Elitzur, {\em Impossibility of Spontaneously Breaking Local Symmetries}, Phys. Rev. D12, 3978–3982 (1975).
\bibitem{Molin} L. Molinari, {\em The Haar Measure of a Lie Group: A Simple Construction},  http://wwwteor.mi.infn.it/~molinari/NOTES/haar.pdf (2019).
\bibitem{Simon2} B. Simon, {\em Representations of Finite and Compact Groups}, American Mathematical Society, Providence, 1996.
\bibitem{Bal} T. Balaban, {\em Ultraviolet Stability of Three-Dimensional Lattice Pure Gauge Field Theories}, Commun. Math. Phys. {102}, 255-275 (1985).
\bibitem {Bal2} T. Balaban, {\em Large Field Renormalization II. Localization, Exponentiation, and Bounds for the $R$ Operation}, Commun. Math. Phys. {122}, 355-392 (1989).
\bibitem{Dim}  J. Dimock, {\em Nonperturbative Renormalization of Scalar Quantum Electrodynamics in $d=3$}, J. Math. Phys. 56, 102304 (2015).
\bibitem{Dim2} {\em Ultraviolet Regularity of QED in $d=3$}, J. Math. Phys. {59}, 012301 (2018).
\bibitem{Dim3} {\em Ultraviolet Stability for QED in $d=3$}, Ann. Henri Poincar\'e {23}, 2113–2205 (2022).
\bibitem{Dim4} {\em Stability for QED in $d=3$: An Overview},  J. Math. Phys. {63}, 042305 (2022).
\bibitem{Nelson}  T.J. Nelson, {\em A Set of Harmonic Functions for the Group ${\mathrm SU}(3)$ as Specialized Matrix Elements of a General Finite Transformation}, J. Math. Phys. 8, 857 (1967).
\bibitem{Holland}  D.F. Holland, {\em Finite Transformations of ${\mathrm SU}(3)$}, J. Math. Phys. 10, 531 (1969).
\bibitem{Holland2}  D.F. Holland, {\em Finite Transformations and Basis States of ${\mathrm SU}(N)$}, J. Math. Phys. 10, 1903 (1969).
\bibitem{Marinov}M.S. Marinov, {\em Invariant Volumes of Compact Groups}, J. Phys. A: Math. Gen. 13, 3357 (1980).
\bibitem{Marinov2}M.S. Marinov, {\em Correction to: Invariant Volumes of Compact Groups}, J. Phys. A: Math. Gen. 14, 543 (1981).
\bibitem{Euler}M. Byrd, {\em The Geometry of ${\mathrm SU}(3)$}, LANL ePrint physics-9708015, (1997).
\bibitem{Euler2} M. Byrd and E.C.G. Sudarshan, {${\mathrm SU}(N)$ Revisited}, J. Phys. A: Math. Gen. 31, 9255 (1998).
\bibitem{Euler3} T. Tilma and E.C.G. Sudarshan, {\em Generalized Euler Angle Parametrization for ${\mathrm SU}(N)$}, Journal of Physics A: Math. Gen. 35, 10467 (2002).
\bibitem{EULER} S.L. Cacciatori and A. Scotti, {\em Compact Lie Groups, Generalised Euler Angles, and Applications}, Universe 2022,8, 492.
\bibitem{Sternberg2} S. Sternberg, {\em Lectures on Differential Geometry}, AMS Chelsea Publishing, vol. 316, 2nd. ed. (1999).
\bibitem{Haber}H.E. Haber, {\em Useful relations among the generators in the defining and adjoint representations of ${\mathrm SU}(N)$}, Sci. Post. Phys. Lecture Notes 21, 1-11 (2021).
\end{thebibliography}
\end{document}